\documentclass[twocolumn]{aastex63}
\usepackage{graphicx}
\usepackage{amsmath}
\usepackage{amssymb}
\usepackage{hyperref}

\def\gtrsim{\mathrel{\hbox{\rlap{\hbox{\lower4pt\hbox{$\sim$}}}\hbox{$>$}}}}
\def\lesssim{\mathrel{\hbox{\rlap{\hbox{\lower4pt\hbox{$\sim$}}}\hbox{$<$}}}}
\def\gtrsim{\mathrel{\hbox{\rlap{\hbox{\lower4pt\hbox{$\sim$}}}\hbox{$>$}}}}
\def\farcs{\hbox{$.\!\!^{\prime\prime}$}}
\def\farcm{\hbox{$.\!\!^{\prime}$}}

\begin{document}

\def\chan{{\sl CXO\ }}

\title{{\sl Chandra} Monitoring of the J1809--1917 Pulsar Wind Nebula and Its Field}

\author{Noel Klingler}
\affil{Department of Astronomy \& Astrophysics, The Pennsylvania State University, 525 Davey Laboratory, University Park, PA, 16802, USA}
\author{Hui Yang}
\affil{Department of Physics, The George Washington University, 725 21st St NW, Washington, DC, 20052, USA}
\author{Jeremy Hare}
\affil{NASA Goddard Space Flight Center, Greenbelt MD, 20771, USA}
\affil{NASA Postdoctoral Program Fellow}
\author{Oleg Kargaltsev}
\affil{Department of Physics, The George Washington University, 725 21st St NW, Washington, DC, 20052, USA}
\author{George G.\ Pavlov}
\affil{Department of Astronomy \& Astrophysics, The Pennsylvania State University, 525 Davey Laboratory, University Park, PA, 16802, USA}
\author{Bettina Posselt}
\affil{Department of Physics, Oxford University, Clarendon Laboratory, Parks Road, Oxford OX1 3PU, UK}

\begin{abstract}
PSR J1809--1917 is a young ($\tau=51$ kyr) energetic ($\dot{E}=1.8\times10^{36}$ erg s$^{-1}$) radio pulsar powering an X-ray pulsar wind nebula (PWN) that exhibits morphological variability. 
We report on the results of a new monitoring campaign by the {\sl Chandra X-ray Observatory (CXO)}, carried out across 6 epochs with a $\sim$7-week cadence.
The compact nebula can be interpreted as a jet-dominated outflow along the pulsar's spin axis. 
Its variability can be the result of Doppler boosting in the kinked jet whose shape changes with time (akin to the Vela pulsar jet).
The deep X-ray image, composed of 405 ks of new and 131 ks of archival {\sl CXO} data, reveals an arcminute-scale extended nebula (EN) whose axis of symmetry aligns with both the axis of the compact nebula and the direction toward the peak of the nearby TeV source HESS J1809--193.
The EN's morphology and extent suggest that the pulsar is likely moving through the ambient medium at a transonic velocity.
We also resolved a faint 7$'$-long nonthermal collimated structure protruding from the PWN.
It is possibly another instance of a ``misaligned outflow'' (also known as a ``kinetic jet'') produced by high-energy particles escaping the PWN's confinement and tracing the interstellar magnetic field lines.
Finally, taking advantage of the 536 ks exposure, we analyzed the point sources in the J1809 field and classified them using multiwavelength data.
None of the classified sources in the field can reasonably be expected to produce the extended TeV flux in the region, suggesting that PSR J1809--1917 is indeed the counterpart to HESS/eHWC J1809--193.
\end{abstract}


\section{INTRODUCTION}
Pulsars are among the most powerful particle accelerators, capable of producing particles up to PeV energies. 
As a neutron star rotates, its rotational energy is imparted to an ultra-relativistic magnetized particle wind in the magnetosphere. 
Outside of the magnetosphere, the wind particles pass through a termination shock (TS), beyond which the gyration of particles in the magnetic field produces synchrotron radiation from radio to $\gamma$-rays, and upscatters photons to produce inverse Compton (IC) radiation from GeV to TeV $\gamma$-rays, which we can see as a pulsar wind nebula (PWN; see, e.g., \citealt{Gaensler2006,Kargaltsev2008,Kargaltsev2015}).

An important factor that affects the morphology of a PWN is the pulsar's Mach number: the ratio of the pulsar's speed to the speed of sound in the ambient medium, $\mathcal{M}\equiv v_{\rm PSR} / c_{\rm amb}$.
The supernova explosions which produce pulsars typically impart them with kick velocities on the order of a few hundred km s$^{-1}$ \citep{Verbunt2017}.
Inside the hot SNR interiors where the sound speed can be several hundreds of km s$^{-1}$, pulsars are usually subsonic ($\mathcal{M}<1$) or transonic ($\mathcal{M}\sim1$), and their anisotropic winds can form polar and equatorial outflows (e.g., jets and torii).
After a few tens of kiloyears, pulsars exit their SNRs and enter the interstellar medium (ISM), where they are often highly supersonic ($\mathcal{M}\gg1$), as ISM sound speeds are typically on the order of a few to a few tens of km s$^{-1}$.
In supersonic PWNe (SPWNe), the ram pressure exerted by the ISM produces a bow shock which confines the pulsar wind to a compact PWN head and pulsar tails which can extend up to a few parsecs behind the pulsar (see, e.g., \citealt{Kargaltsev2017,Reynolds2017}).
At high Mach numbers, the PWN structures (jets and torii) will be crushed by the ram pressure of the oncoming medium and mixed together (e.g., PSR J1101--6101; $v\sim1000$ km s$^{-1}$; \citealt{Pavan2014,Pavan2016}).
For pulsars whose velocities are substantially supersonic, the stand-off distance of the bow shock apex can be approximated as $r_s = (\dot{E}/4\pi c \rho v_{\rm PSR}^2)^{1/2}$ (where $\dot{E}$ is the pulsar's spin-down power, and $\rho$ is the ISM density). 
Such distances are usually on the order of a few arcsec (for typical pulsar distances of a few kpc).
For transonic PWNe ($\mathcal{M}\sim1$), the jets and torus will be only mildly perturbed, and emission can be seen much further ahead of the pulsar (usually about an arcminute; see, e.g., PSR B1706--44 and PSR J2021+3651; \citealt{Romani2005,vanEtten2008}).
Even if transonic pulsars do not form bow shocks or tails, their motion through the ambient medium can still shape and confine the wind into dome-like shapes.

A few supersonic pulsars have been seen to produce very elongated and collimated outflows seen in X-rays which extend up to parsec-scale distances beyond the boundaries of their bow shocks.
These ``misaligned outflows'' are puzzling in that they can be strongly offset from their pulsars' direction of motion, and because they appear to be unaffected by the ram pressure which confines most of the pulsar wind to the direction behind the pulsars.
Misaligned outflows have been seen in a handful of SPWNe:  B2224+65 (the Guitar PWN; \citealt{Hui2007}), J1101--6101 (the Lighthouse PWN; \citealt{Pavan2014,Pavan2016}), J1509--5850 \citep{Klingler2016a}, B0355+54 \citep{Klingler2016b}, J2055+2539 \citep{Marelli2016,Marelli2019}, and PSR J2030+4415 \citep{deVries2020}, and some transonic PWNe: J2021+3651 (the Dragonfly PWN; \citealt{vanEtten2008}), G3271--1.1 (the Snail PWN; \citealt{Temim2009}), and G291.0--0.1 (MSH 11--62; \citealt{Slane2012}).
To explain the misaligned outflow produced by the Guitar PWN, \citet{Bandiera2008} suggested that these outflows can be formed if the gyroradii of high-energy electrons exceed the bow shock stand-off distance (which is particularly small in SPWNe).
In this scenario, the electrons can not be contained within the extent of the bow shock, and can leak into the ISM, where they then travel along and trace the ISM magnetic fields.
Thus, the misalignment with respect to the pulsar's direction of motion is due to the direction of the local ambient magnetic field being different from the direction of motion.
Simulations of SPWNe by \citet{Barkov2019} and \citet{Olmi2019} show that these misaligned outflows (also referred to as ``kinetic jets''; not to be confused with jets along a pulsar's spin axis) can also be produced when the ISM magnetic field lines reconnect with the PWN magnetic field lines, allowing some particles to escape.
Magnetic bottles can occur at the reconnection site, which can filter out low energy particles, preventing them from escaping.
Both of these hypotheses (particle leakage via large gyroradii, and escape along reconnected magnetic field lines) are consistent with the lack of detection of these outflows at energies below the X-ray range (even when their associated PWNe are seen in radio), and with the similarly hard spectra these outflows all exhibit ($\Gamma\approx1.6-1.7$).
The asymmetry (one-sidedness) of the outflow has been suggested to be the result of the ISM magnetic field lines being able to reconnect with PWN magnetic field lines on only one side of the pulsar's equatorial plane, since each side will have opposite directions of field lines (see Figures 4 and 8 of \citealt{Barkov2019}).

PSR J1809--1917 (J1809 hereafter) is an energetic young 82.7 ms radio pulsar with spin-down power $\dot{E}=1.8\times10^{36}$ erg s$^{-1}$ and characteristic spin-down age $\tau_{\rm sd}=51$ kyr (more parameters listed in Table \ref{tbl-parameters}).
It is located in the Galactic plane ($l = 11.09^\circ$, $b=0.08^\circ$) at a dispersion measure distance $d=3.3$ kpc (when using the \citealt{YMW2017} free electron density model); we adopt this distance below. 
PSR J1809 is not associated with any known SNR (see \citealt{Castelletti2016,Klingler2018a}), suggesting it has either left its parent SNR and/or that its parent SNR is too faded, cooled, and dissipated to be seen in radio or X-rays.
The pulsar position is projected within the extent of the TeV sources HESS J1809--193 (which has been suggested to be associated with the pulsar/PWN; \citealt{Aharonian2007,Kargaltsev2007,HESS2018,Klingler2018a}) and eHWC J1809--193 (see Table 1 of \citealt{HAWC2019}).

\citet{Kargaltsev2007} reported an observation of the field of J1809 with the {\sl Chandra X-ray Observatory (CXO)} in 2004. 
In addition to the pulsar, they detected a compact ($\sim$10$''$) PWN, as well as faint emission extending $1'-2'$ from the pulsar.
This target was again observed with the {\sl CXO} in 2013 and 2014. 
Klingler et al.\ (2018a; hereafter K+18) reported the analysis of these two observations and the first observation of 2004. 
K+18 found that the bright compact nebula (CN; $\Gamma_{\rm CN}=1.55\pm0.09$) displayed morphological and flux variability on timescales of months to years.
Small (arcsecond) scale elongated emission with a hard spectrum ($\Gamma=1.2\pm0.1$) was seen in the pulsar's vicinity, leading K+18 to hypothesize the variable CN may be associated with a pulsar jet (i.e., a collimated outflow launched along the pulsar spin axis).
They also found that the fainter arcminute-scale extended nebula (EN) exhibited the same spectrum as the CN and shared the axis of symmetry with the CN, which is aligned with the direction to the brightest part of HESS J1809.

In this paper we report on the results of 9-month {\sl Chandra} monitoring campaign (consisting of 6 epochs spaced roughly 7 weeks apart; PI: Pavlov) carried out to study the variability in the CN.
Since this field has one of the deepest {\sl CXO} exposures taken in the Galactic Plane, we have also analyzed the field X-ray source content and investigated whether there are other X-ray sources that could contribute to the TeV emission of the extended source HESS J1809--193.
The paper is structured as follows.
In Section 2 we list the observations and data reduction details.
In Section 3 we present our results, and in Section 4 we discuss their implications.
In Section 5 we present an analysis of the point sources found in J1809 field using the combined 536 ks exposure.
In Section 6 we summarize our results.

\begin{deluxetable}{lc}
\tablecolumns{9}
\tablecaption{Observed and Derived Pulsar Parameters \label{tbl-parameters}}
\tablewidth{0pt}
\tablehead{\colhead{Parameter} & \colhead{Value} }
\startdata
R.A. (J2000.0) & 18 09 43.147(7)  \\
Decl. (J2000.0) & --19 17 38.1(13)  \\
Epoch of position (MJD) & 51506  \\
Galactic longitude (deg) & 11.09  \\
Galactic latitude (deg) & --0.08  \\
Spin period, $P$ (ms) & 82.7  \\
Period derivative, $\dot{P}$ (10$^{-14}$) & 2.553 \\
Dispersion measure, DM (pc cm$^{-3}$) & 197.1  \\
Distance, $d$ (kpc) & 3.3, 3.5  \\
Surface magnetic field, $B_s$ (10$^{12}$ G) & 1.5  \\
Spin-down power, $\dot{E}$ (10$^{36}$ erg s$^{-1}$) & 1.8  \\
Spin-down age, $\tau_{\rm sd} = P/(2\dot{P})$ (kyr) & 51.3 
\enddata
\tablenotetext{}{Parameters are from the ATNF Pulsar Catalog \citep{Manchester2005}.  The DM distance estimates listed correspond to those obtained using the Galactic free electron density models of Yao, Manchester, \& Wang (2017), and for \citet{Cordes2002}.}
\end{deluxetable}

\section{OBSERVATIONS AND DATA REDUCTION}
Fourteen observations of PSR J1809 (405 ks) were carried out over a span of roughly 9 months.
The observations were grouped into 6 ``new'' epochs ($\approx$70 ks each) spaced apart by 6-8 week intervals in order to monitor morphological changes in the nebula.
For all observations (new and archival), the ACIS detector was operated in Very Faint timed exposure mode (3.24 s time resolution).
ObsID 3853 was taken with the ACIS-S, while all other observations were taken with the ACIS-I.
We list details of the observations in Table \ref{tbl-obs}.

\begin{deluxetable}{ccccc}
\tablecolumns{4}
\tablecaption{17 \chan Observations of J1809 field\label{tbl-obs}}
\tablewidth{0pt}
\tablehead{
\colhead{ObsID} & \colhead{Exposure,} & \colhead{Date} & \colhead{Epoch} & \colhead{$\theta$,} \\ 
\colhead{} & \colhead{ks} & \colhead{} & \colhead{} & \colhead{arcmin} }
\startdata
3853 & 19.70 & 2004 Jul 21 & 1 & $0\farcm7$ \\
\hline
14820 & 46.75 & 2013 Sep 29 & 2 & $0\farcm7$ \\
\hline
16489 & 64.86 & 2014 May 25 & 3 & $1\farcm3$ \\
\hline
20327 & 15.32 & 2018 Feb 08 & 4 & $1\farcm0$ \\
20962 & 20.93 & 2018 Feb 09 & 4 & $1\farcm0$ \\
20967 & 15.05 & 2018 Feb 10 & 4 & $1\farcm9$ \\
20963 & 16.33 & 2018 Feb 12 & 4 & $1\farcm0$ \\
\hline
20328 & 33.60 & 2018 Apr 02 & 5 & $1\farcm0$ \\
21067 & 30.64 & 2018 Apr 03 & 5 & $1\farcm1$ \\
\hline
20329 & 29.66 & 2018 May 25 & 6 & $1\farcm0$ \\
21098 & 39.37 & 2018 May 27 & 6 & $1\farcm1$ \\
\hline
20330 & 32.62 & 2018 Jul 18 & 7 & $0\farcm6$ \\
21126 & 40.01 & 2018 Jul 19 & 7 & $0\farcm6$ \\
\hline
20331 & 39.53 & 2018 Aug 30 & 8 & $0\farcm6$ \\
21724 & 25.71 & 2018 Sep 01 & 8 & $0\farcm5$ \\
\hline
20332 & 37.58 & 2018 Nov 03 & 9 & $0\farcm8$ \\
21874 & 28.67 & 2018 Nov 04 & 9 & $0\farcm7$
\enddata
\tablenotetext{}{Note -- $\theta$ is the angular distance between the pulsar and the optical axis of the telescope.  All exposures were taken using the ACIS-I detector, except for ObsID 3853 which was taken with ACIS-S.}
\end{deluxetable}

The data were reprocessed using the {\sl Chandra} Interactive Analysis of Observations (CIAO) software package version 4.11, using the Calibration Data Base (CALDB) version 4.8.3.
We used the standard {\tt chandra\_repro} pipeline to apply the latest calibrations.

We removed the field point sources (detected at $\geq3\sigma$ by {\tt wavdetect}; \citealt{Freeman2002}) from the exposure-map-corrected images\footnote{See \url{https://cxc.cfa.harvard.edu/ciao/threads/diffuse_emission/} for details.}, and combined them all using {\tt merge\_obs}, in order to resolve and study the faint extended features of the PWN.
All images and spectra were restricted to the 0.5--8 keV energy range unless otherwise specified.
In all images, North is up, and East is left.

\begin{figure}
\includegraphics[width=0.98\hsize,angle=0]{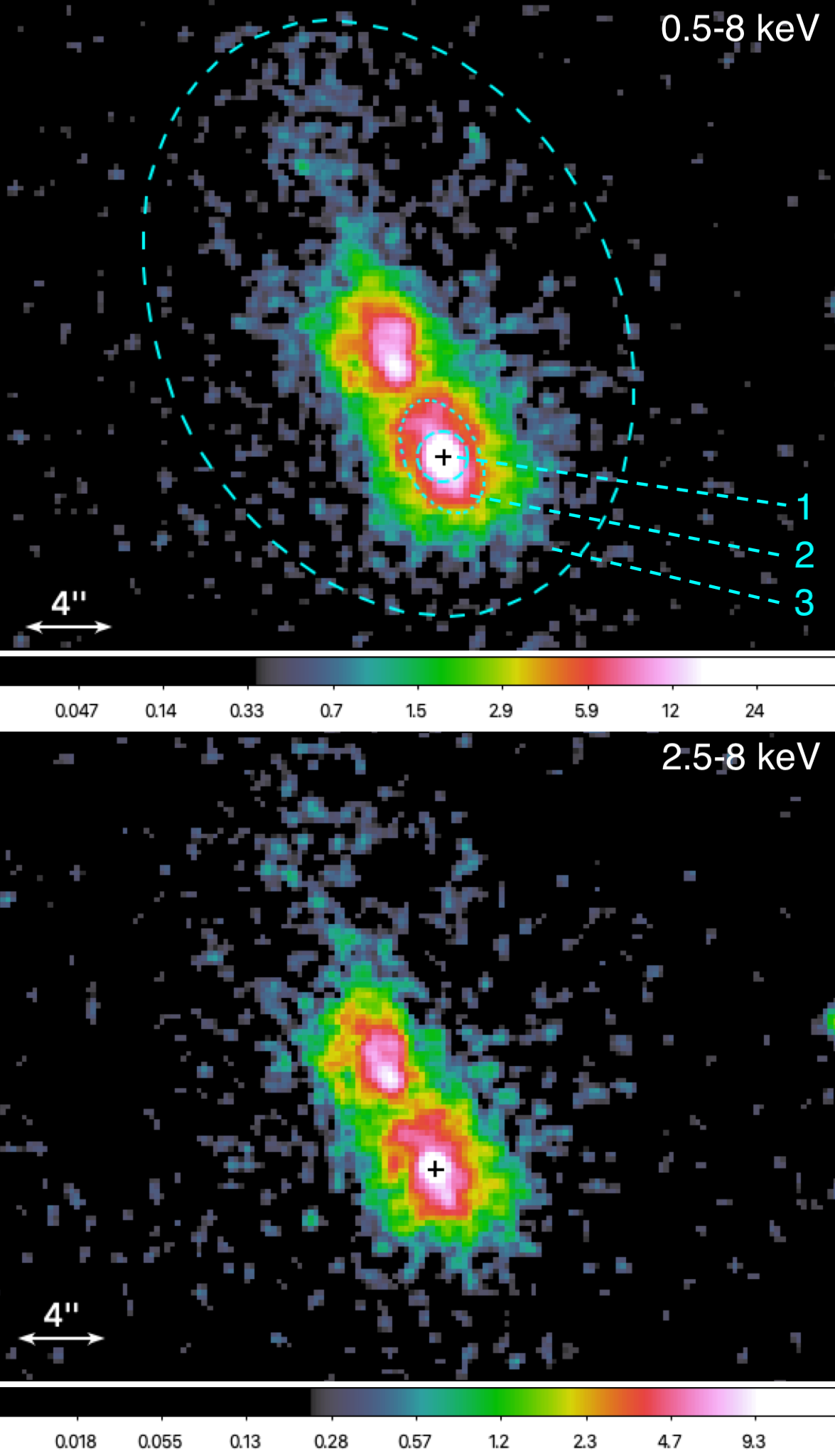}
\caption{{\sl Chandra} ACIS images of the CN, made by merging all existing observations (binned by a factor of 0.5; i.e., pixel size of $0\farcs246$; and smoothed with an $r=1\farcs5$ Gaussian kernel).  Top: 0.5--8 keV; bottom: 2.5--8 keV (we select this energy to allow a comparison with Figure 3 of K+18).  The black cross marks the pulsar position.  The following regions are shown in the top panel:  1 -- pulsar (the $r=1\farcs25$ circle); 2 -- pulsar vicinity (the area enclosed within the smaller ellipse, excluding region 1); 3 -- CN (the area enclosed within the larger ellipse, excluding regions 1 and 2).  The color bar is in units of counts pixel$^{-1}$.}
\label{fig-cn-merged}
\end{figure}

To obtain the absorbing Hydrogen column density, $N_{\rm H}$, we initially fitted the spectra from the compact nebula (CN) region (region 3 in Figure \ref{fig-cn-merged}) from all observations simultaneously, since it is small enough that synchrotron cooling effects across its extent should be minimal, and since no significant spectral variability between observations was seen (see Section 3).
Our best-fit value, $N_{\rm H}=(0.69\pm0.08)\times10^{22}$ cm$^{-2}$, is consistent with that previously reported by K+18 and \citet{Kargaltsev2007}, $N_{\rm H} = 0.7\times10^{22}$ cm$^{-2}$;  in all spectral fits discussed below we set $N_{\rm H}$ to $0.7\times10^{22}$ cm$^{-2}$.
Fitting the spectra from the CN surroundings and extended nebula (EN) with a power-law (PL) fit yielded similar values, $N_{\rm H} = (0.73\pm0.16)\times10^{22}$ and $(0.78\pm0.11)\times10^{22}$ cm$^{-2}$, respectively.

\section{RESULTS}

\subsection{Pulsar}

\subsubsection{Proper Motion}

In order to measure the pulsar's proper motion, we matched the observations to a common astrometric frame using an external catalog, as outlined in the corresponding CIAO thread\footnote{See \url{https://cxc.harvard.edu/ciao/threads/reproject\_aspect/}}.
The frame was defined by the {\sl Gaia} DR2 catalog\footnote{{\sl Gaia} DR2 data were collected between 25 July 2014 (10:30 UTC) and 23 May 2016 (11:35 UTC).  The reference epoch is J2015.5.  See \url{http://www.cosmos.esa.int/web/gaia/dr2 for details.}} \citep{Gaia2018}. 
We cross-correlated the positions of the {\sl Gaia} DR2 stars with the X-ray sources detected by {\tt wavdetect} in each of the {\sl CXO} observations\footnote{If an X-ray source is variable, it may not be  present in all {\sl CXO} observations but still can be used for alignment if it is present in more than one observation.}.
We excluded the pulsar and any sources within a $30''$ radius of it to prevent bright nebular emission from being misidentified as point sources.
We also required the selected X-ray sources to be detected with signal-to-noise ratios of $\ge3$ and be within $5'$ of the optical axis.
After cross-correlation with the {\sl Gaia} DR2 catalog we selected X-ray sources with {\sl Gaia} counterparts within a $1''$ radius from the X-ray source position determined by {\tt wavdetect}.  
Each observation had at least 3 cross-correlated reference sources.
We then used CIAO's {\tt wcs\_match} to calculate the translation transformation to shift the input (X-ray) sources to the reference ({\sl Gaia} DR2) source locations in a way that minimizes the differences in their positional offsets.  
We iterated the transformation calculation process until the largest source pair position error became less than $0\farcs5$ by iteratively rejecting those {\sl Chandra-Gaia} pairs that violated this condition. 
The $0\farcs5$ was criterion chosen because it is approximately the typical astrometric accuracy of {\sl Chandra}.
For each observation we then calculated the root mean square of the offsets for the pairs of sources that were used in the final transformation.
Following the CIAO thread mentioned above, we then used {\tt wcs\_update} to recalculate the aspect solutions and update the world coordinate system (WCS) parameters in the WCS transformation blocks of the Level 2 event list files. 
    
\begin{figure*}
\includegraphics[width=0.98\hsize,angle=0]{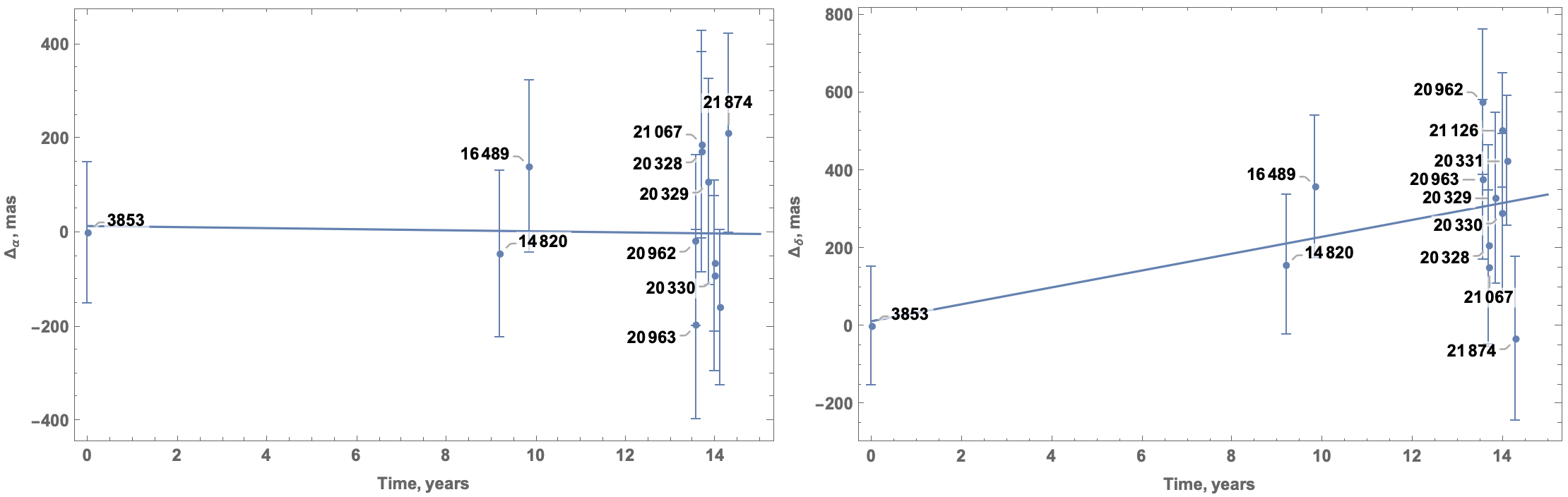}
\caption{Pulsar proper motion in R.A.\ (left) and decl.\ (right).  The lines represent the best fits to the data, $\mu_{\alpha}=-1\pm9$ mas yr$^{-1}$ and $\mu_{\delta}=24\pm11$ mas yr$^{-1}$.  The error bars shown represent 1$\sigma$ uncertainties.}
\label{fig-proper-motion}
\end{figure*}

After the observations were aligned to the common reference frame (based on {\sl Gaia} DR2), we measured the position of the pulsar centroid by calculating the average R.A.\ and decl.\ position of all counts within $1\farcs5$ of the brightest pixel (in the vicinity of the pulsar) in the images  binned by a factor of 0.5 and smoothed with an $r=3''$ Gaussian kernel.
The uncertainties of the pulsar positions were calculated using CIAO's {\tt celldetect} tool\footnote{See \url{http://cxc.harvard.edu/ciao/download/doc/detect\_manual/wav\_theory.html}}.  
The positions and uncertainties are also given in Table \ref{tbl-psr-coords}. (We did not include ObsIDs 20327, 20967, 21098, 21724, and 20332 in the proper motion measurement either because the astrometric correction did not converge to the required precision or because the pulsar image appeared to be distorted.) 
The total uncertainty of the pulsar position also includes the uncertainty of alignment for the each observation to the {\sl Gaia} DR2 reference frame, reported by {\tt wcs\_match}\footnote{ {\tt wcs\_match} reports the average residuals after the final iteration, which we take as the uncertainty of alignment,}, which was added in quadrature to the pulsar centroiding uncertainty. 
The coordinates of the pulsar (in R.A.\ and decl.) and their uncertainties for each observation epoch are shown in Figure \ref{fig-proper-motion}.  
We fitted a linear function to the pulsar coordinates over time (the best fits are shown in the corresponding panels) and found the proper motion $\mu_{\alpha}=-1\pm9$ mas yr$^{-1}$ and $\mu_{\delta}=24\pm11$ mas yr$^{-1}$. 
The net result is that there is an indication of proper motion northward, but only at a 95\% confidence level.

\begin{deluxetable}{cccccc}
\tablecolumns{6}
\tablecaption{Pulsar Coordinates in 12 {\sl CXO} Observations Used for the Proper Motion Measurement\label{tbl-psr-coords}}
\tablewidth{0pt}
\tablehead{
\colhead{ObsID} & \colhead{Exposure} & \colhead{R.A.} & \colhead{Decl.} & \colhead{$\sigma_\alpha$} & \colhead{$\sigma_\delta$} \\
\colhead{} & \colhead{s} & \colhead{deg} & \colhead{deg} & \colhead{mas} & \colhead{mas} }
\startdata
 3853 & 19702 & 272.429702 & $-$19.293967 & 150 & 150 \\
14820 & 46746 & 272.429690 & $-$19.293923 & 180 & 180 \\ 
16489 & 64864 & 272.429741 & $-$19.293867 & 180 & 180 \\ 
20328 & 33604 & 272.429750 & $-$19.293909 & 260 & 260 \\ 
20329 & 29661 & 272.429732 & $-$19.293875 & 220 & 220 \\ 
20330 & 32617 & 272.429677 & $-$19.293886 & 200 & 200 \\ 
20331 & 39531 & 272.429658 & $-$19.293849 & 170 & 170 \\ 
20962 & 20926 & 272.429698 & $-$19.293807 & 180 & 190 \\ 
20963 & 16334 & 272.429648 & $-$19.293862 & 200 & 210 \\ 
21067 & 30643 & 272.429754 & $-$19.293925 & 200 & 200 \\ 
21126 & 40008 & 272.429684 & $-$19.293827 & 140 & 150 \\ 
21874 & 28667 & 272.429761 & $-$19.293976 & 210 & 210    
\enddata
\tablenotetext{}{The coordinates correspond to those after the alignment with the {\sl Gaia} DR2 catalog was performed.  The uncertainties $\sigma_\alpha$ and $\sigma_\delta$, are 1$\sigma$.}
\end{deluxetable}

\subsubsection{Spectrum}
We extracted the pulsar spectrum from each new observation (epochs 4--9) using a $1\farcs2$ aperture radius, and then combined the spectra and response files using the CIAO tool {\tt combine\_spec} since the pulsar was positioned at approximately the same off-axis angle in all of the new observations.
A minimum of 20 counts was required for each spectral energy bin.

We first tried fitting the spectrum with a simple absorbed PL model. 
We obtained $\Gamma=1.70\pm0.07$ and a reduced $\chi^2_{\nu}=1.44$ (for $\nu=44$ d.o.f.), with systematic residuals at both low and high energies (i.e., the data exceed the model at $<$1.5 keV, $>$5 keV). 
An absorbed blackbody (BB) model (XSPEC's ``bbodyrad'') provided an even worse fit, with a reduced $\chi^2_{44} = 4.3$.  Thus, a single-component model (PL or BB) fit to the pulsar spectrum is not satisfactory.

Next, we tried fitting the spectrum with a two-component PL model to account for possible contamination from the surrounding nebula, setting $\Gamma_{\rm PWN}=1.23$ (the best-fit value for the emission in the pulsar vicinity; see below).
Although the fit yielded a satisfactory $\chi^2_{43}=1.00$, we obtained an unusually large $\Gamma_{\rm PSR}=4.62\pm0.72$.

We then tried fitting the spectrum with a PL + BB model, again setting $\Gamma=1.23$. 
We obtained $kT=0.20\pm0.02$ keV (corresponding to a temperature $T = 2.3\pm0.3$ MK), and a bbodyrad normalization of $2.0_{-1.2}^{+2.8}$ (corresponding to an emitting blackbody radius $r=460_{-430}^{+260}$ m, at 3.3 kpc), for $\chi^2_{43}=1.005$.
The PL normalization was $\mathcal{N}=(6.2\pm0.3)\times10^{-6}$ photons s$^{-1}$ cm$^{-2}$ keV$^{-1}$ at 1 keV.
This fit corresponds to an absorbed flux of $(4.8_{-0.3}^{-0.1})\times10^{-14}$ erg cm$^{-2}$ s$^{-1}$ (0.5--8 keV).

Allowing $\Gamma$ to be a free parameter, we obtained $\Gamma=1.28\pm0.15$ (consistent with the emission from region 2); see K+18, and below), $\mathcal{N}=(6.6\pm0.1)\times10^{-6}$ photons s$^{-1}$ cm$^{-2}$ keV$^{-1}$ at 1 keV, $kT=0.19\pm0.03$ keV (corresponding to $T=2.2\pm0.4$ MK), and a bbodyrad normalization of $2.4_{-1.6}^{+5.9}$ (corresponding to $r=530_{-200}^{+430}$ m, at 3.3 kpc)
with $\chi^2_{42}=1.027$.
This fit yields an absorbed flux of $(4.7_{-0.4}^{+0.1})\times10^{-14}$ erg cm$^{-2}$ s$^{-1}$.
The PL+BB model is the most plausible, as it implies the presence of a hot polar cap (or caps) whose temperature and size are consistent with those seen in other pulsars of similar spin-down power and age.
These parameters are comparable to the best-fit model parameters obtained by K+18 ($\Gamma=1.2$, $kT=0.17 \pm 0.02$ keV, and a normalization corresponding to an emitting blackbody radius of $840^{+410}_{-260}$ m).

For completeness, we also tried fitting the pulsar spectrum with a PL ($\Gamma_{\rm PWN}=1.23$) + PL ($\Gamma_{\rm PSR}$) + BB model as we did in K+18.
We obtained a poorly-constrained $\Gamma_{\rm PSR}=3.7\pm7.2$, and $kT=0.17\pm0.06$ keV (with a bbodyrad normalization $2.7_{-2.7}^{+10}$).
Although the fit is satisfactory with $\chi^2_{41}=1.041$, this more complex model is not required by the data.

\subsection{Compact Nebula (CN)}

\subsubsection{Morphology}

In Figure \ref{fig-cn-merged} we present a merged image of the CN, made by combining all observations.
The CN still appears elongated in the NE-SW direction, with the brightest part occupying an elliptical region with major/minor axes of $16''$/$8''$ (and $28''$/$20''$ for the full extent; enclosed by region 3).
Small scale ($<$3$''$) collimated emission still appears extending outward from the pulsar (region 2), along the same line as that reported by K+18, who interpreted these as evidence of small-scale pulsar jets. 

The new observations (epochs 4--9) reveal the presence of a prominent $\approx$3$''$ diameter ``blob'' located about 4$''$ NE of the pulsar.

\subsubsection{Spectrum}
We use the same regions and numbering as defined in K+18, with the modification of region 3 (the CN) from a contour into a larger ellipse to better accommodate the morphological variability of the CN across all epochs (see Figure \ref{fig-cn-merged}).
For region 2 we obtained a hard spectrum with $\Gamma_2=1.23\pm0.09$, which is consistent with the previously reported result.
This spectrum is also consistent with the PL component of the spectrum extracted from the pulsar vicinity (region 1; $\Gamma=1.28\pm0.15$; see above).
This suggests that the PL component seen in the spectrum extracted from the pulsar region may be due to contamination from the surrounding nebula, or that both the emission from the pulsar region and its surrounding vicinity have the same spectrum.
We see evidence of spectral softening between this region and the CN's spectrum (region 3), with $\Gamma_3=1.52\pm0.03$.
These results are similar to those reported by K+18 ($\Gamma_{\rm 2,old}=1.20\pm0.11$, and $\Gamma_{\rm 3,old}=1.53\pm0.08$).
We list all spectral fit results (for these regions and the ones to be discussed subsequently) in Table \ref{table-pwn-spectra}.

\begin{deluxetable*}{cccccccccc}
\tablecolumns{4}
\tablecaption{Spectral Fit Results for PWN Regions}
\tablewidth{0pt}
\tablehead{\colhead{Region} & \colhead{Name} & \colhead{Area} & \colhead{Counts} & \colhead{$\mathcal{B}$} & \colhead{$\Gamma$} & \colhead{$\mathcal{N}_{-5}$} & \colhead{$\chi_\nu^2$} & \colhead{$F_{-13}$} & \colhead{$F_{-13}^{\rm unabs}$} }
\startdata
2 & Pulsar Vicinity & 9.44 & $1019\pm32$ & 15 & $1.23\pm0.09$ & $0.55\pm0.05$ & 1.06 (51) & $0.39\pm0.09$ & $0.50\pm0.03$ \\
3 & Compact Nebula & 464 & $5206\pm75$ & 25 & $1.52\pm0.03$ & $3.36\pm0.13$ & 1.29 (192) & $1.62\pm0.03$ & $2.25\pm0.04$ \\
4 & CN Vicinity & 2623 & $1879\pm66$ & 25 & $1.72\pm0.08$ & $1.48\pm0.12$ & 0.92 (140) & $0.55\pm0.03$ & $0.82\pm0.03$ \\
5 & Extended Nebula & 42671 & $9138\pm302$ & 200 & $1.74\pm0.05$ & $7.98-10.35$ & 1.31 (211) & $2.6-4.9$ & $3.9-7.3$ \\
6 & Outflow & 42659 & $4031\pm321$ & 200 & $1.74\pm0.12$ & $2.25-5.45$ & 1.15 (123) & $0.8-1.8$ & $1.2-2.7$
\enddata
\tablenotetext{}{Spectral fit results for the different regions of the PWN.  Listed are the region number (following the numbering of K+18), region name, area (in arcsec$^2$), net counts, minimum number of counts per energy bin $\mathcal{B}$, photon index $\Gamma$, PL normalization $\mathcal{N}_{-5}$ in units of $10^{-5}$ photons s$^{-1}$ cm$^{-2}$ keV$^{-1}$ at 1 keV, reduced $\chi^2_{\nu}$ ($\nu$ d.o.f.), and absorbed and unabsorbed 0.5--8 keV fluxes $F_{-13}$ and $F_{-13}^{\rm unabs}$ in units of $10^{-13}$ erg cm$^{-2}$ s$^{-1}$.
The chip gap crossed through regions 5 and 6 (at varying positions due to varying roll angles), so we left the normalizations (and fluxes) as free parameters and list the range of values they spanned.  
In all fits we set $N_{\rm H}=0.7\times10^{22}$ cm$^{-2}$. }
\label{table-pwn-spectra}
\end{deluxetable*}

\subsubsection{Variability}

\begin{figure*}
\includegraphics[width=0.98\hsize,angle=0]{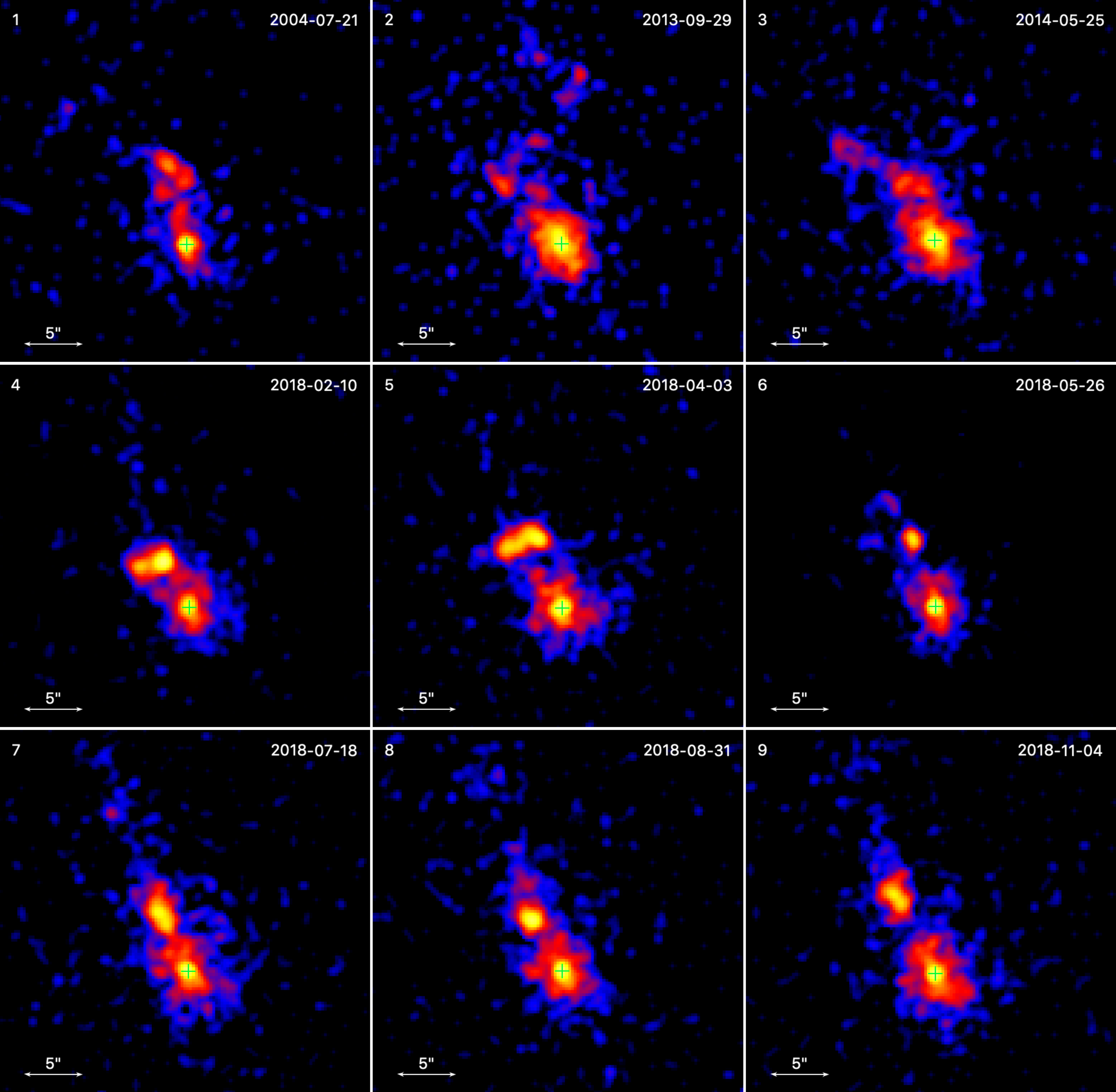}
\caption{{\sl Chandra} ACIS images (0.5--8 keV) of the compact nebula, showing the morphological variability across the 9 observation epochs (the ObsIDs and exposures are given in Table \ref{tbl-obs}).  Images were binned by a factor of 0.5 and smoothed with a 3-pixel ($r=0\farcs75$) Gaussian kernel.  The pulsar position in each image is marked with a green cross.}
\label{fig-variability}
\end{figure*}

In Figure \ref{fig-variability} we present images of the CN from the 6 new epochs (in addition to the three archival images, for comparison).
The new images of J1809 show that the CN continues to exhibit changes over timescales of $\sim$months.
Across epochs 1-3, the variability was best-described as a tail-wagging motion (as was previously reported in K+18).
However in epochs 4-9, the CN emission appears to be dominated by a blob-like structure,  appearing to be 2-4$''$ in size and positioned 4-5$''$ NE of the pulsar, but whose position, shape, and orientation vary between the epochs.

\begin{figure*}
\includegraphics[width=0.98\hsize,angle=0]{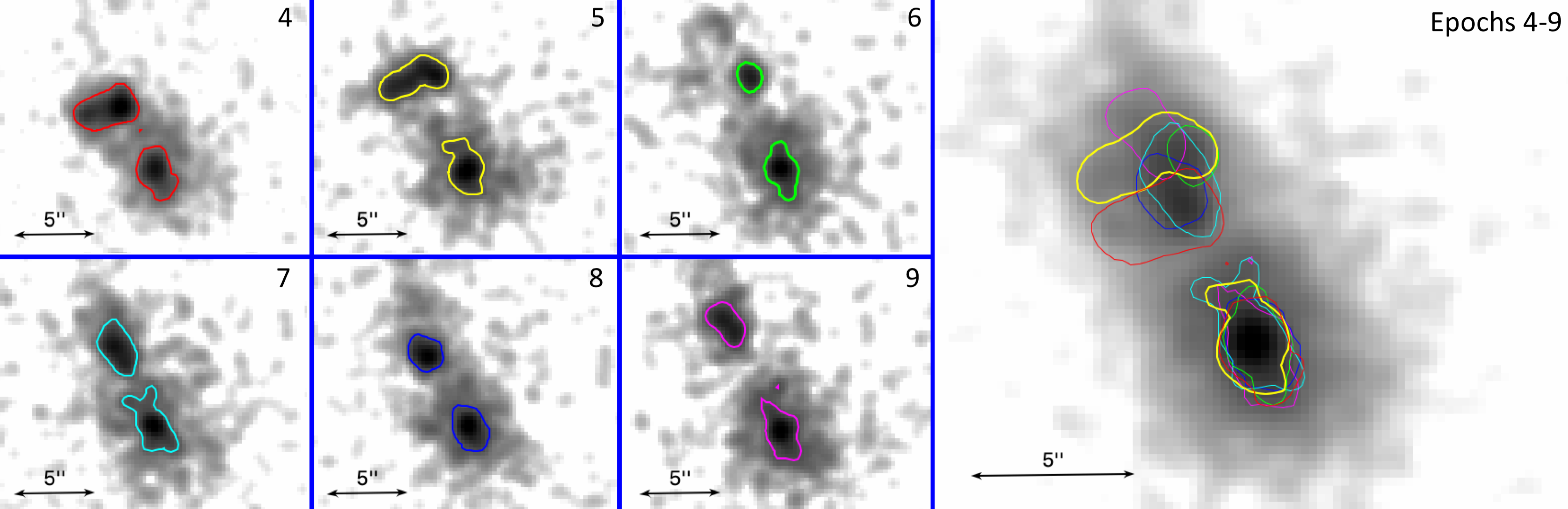}
\caption{{\sl Small panels:} Images of epochs 4--9 (binned by a factor of 0.5, and smoothed with an $r=3$ pixel Gaussian kernel; best viewed in color).  The contours were created using DS9, by setting one contour to a level of 1 count pixel$^{-1}$ and the smoothness option set to 1. {\sl Large panel (right):} Merged image, with all contours overlayed.  In all images North is up, and East is left.  In each image, the pulsar is located within the southern contour.}
\label{fig-blobs-contours}
\end{figure*}

In Figure \ref{fig-blobs-contours} we present contours of the CN across the 6 new epochs to compare the morphological changes over short-term timescales (intervals of 6-8 weeks).
In epochs 4 and 5 the blob appears elongated in the SE-NW direction; i.e., perpendicular to the CN's (and EN's) axis of symmetry.
In epochs 5 and 8 the blob appears round, and in epochs 7 and 9 the blob appears elongated in the NE-SW direction.
When overlaying the contours of the blob from epochs 4-9 in the same frame (as shown in the right panel of Figure \ref{fig-blobs-contours}), there appears to be no evidence of steady motion, either away from or toward the pulsar.

In some epochs, the blob appears comparable in brightness to the pulsar itself (see Figures \ref{fig-variability} and \ref{fig-blobs-contours}). 
To investigate possible spectral variability in the blob, we used an $r=2\farcs5$ extraction radius to obtain the spectra from each of the ``new'' epochs (i.e., epochs 4--9).  
We list the results in Table \ref{table-blobs}; the fit results for epochs 1-3 are not listed as the blob is not present in those observations.
Since the individual observations which comprise epochs 4--9 were adjacent in time (i.e., observations within an epoch were taken within 2 days of each other) and were taken at the same roll and off-axis angles, we combined the spectra and corresponding response files from multiple observations (within the same epoch) using {\tt combine\_spec}.
We found no evidence of any significant spectral variability across the epochs, although the count rates varied significantly.
We then simultaneously fitted the spectra from epochs 4--9 with the PL slopes tied, but with the normalization for each epoch left as free parameters.
For the best-fit $\Gamma=1.34\pm0.06$ (with $\chi^2_{\nu}=0.99$ for $\nu=87$ degrees of freedom), we found that the blob's flux varied significantly by up to a factor of 3 across epochs, ranging from $(12.4_{-0.7}^{+0.4})$ to $(4.2\pm0.4)\times10^{-14}$ erg cm$^{-2}$ s$^{-1}$ (in the 0.5--8 keV band).

\begin{deluxetable*}{ccccccc}
\tablecolumns{4}
\tablecaption{Blob Spectral Analysis}
\tablewidth{0pt}
\tablehead{\colhead{Epoch} & \colhead{Count Rate,} & \colhead{$\Gamma$} & \colhead{$\mathcal{N}_{-5}$,} & \colhead{$\chi^2_\nu$} & \colhead{$F_{-14,\ \Gamma=1.34}$,} \\ \colhead{} & \colhead{counts ks$^{-1}$} & \colhead{} & \colhead{} & \colhead{(dof)} & \colhead{} }
\startdata
4 & $6.20\pm0.30$ & $1.33\pm0.11$ & $2.01\pm0.25$ & 0.55 (17) & $12.4_{-0.7}^{+0.4}$ \\
5 & $4.89\pm0.28$ & $1.46\pm0.14$ & $1.73\pm0.26$ & 1.37 (17) & $9.3\pm0.6$ \\
6 & $1.85\pm0.16$ & $1.48\pm0.24$ & $0.80\pm0.21$ & 0.96 (10) & $4.2\pm0.4$ \\
7 & $3.75\pm0.23$ & $1.14\pm0.16$ & $0.95\pm0.17$ & 1.35 (14) & $7.1_{-0.4}^{+0.5}$ \\
8 & $3.39\pm0.23$ & $1.34\pm0.16$ & $1.09\pm0.20$ & 0.87 (15) & $6.9_{-0.6}^{+0.5}$ \\
9 & $3.18\pm0.22$ & $1.41\pm0.16$ & $1.09\pm0.20$ & 0.81 (14) & $6.2\pm0.4$
\enddata
\tablenotetext{}{For each epoch we fitted the blob spectra extracted from an $r=2\farcs5$ circular aperture.  We list the count rate, photon index $\Gamma$, normalization $\mathcal{N}_{-5}$ (in units of $10^{-5}$ photon s$^{-1}$ cm$^{-2}$ keV$^{-1}$, at 1 keV), and the reduced $\chi^2$ for $\nu$ degrees of freedom.  $F_{-14,\ \Gamma=1.34}$ represents the absorbed flux (in units of $10^{-14}$ erg cm$^{-2}$ s$^{-1}$) when the blobs were simultaneously fitted to the same PL slope ($\Gamma=1.34$, the average value), but with their normalizations left as free parameters.  We used $N_{\rm H}=0.7\times10^{22}$ cm$^{2}$ for all fits.}
\label{table-blobs}
\end{deluxetable*}

\subsection{Extended Nebula (EN)}

\subsubsection{Morphology}
\begin{figure*}
\includegraphics[width=0.98\hsize,angle=0]{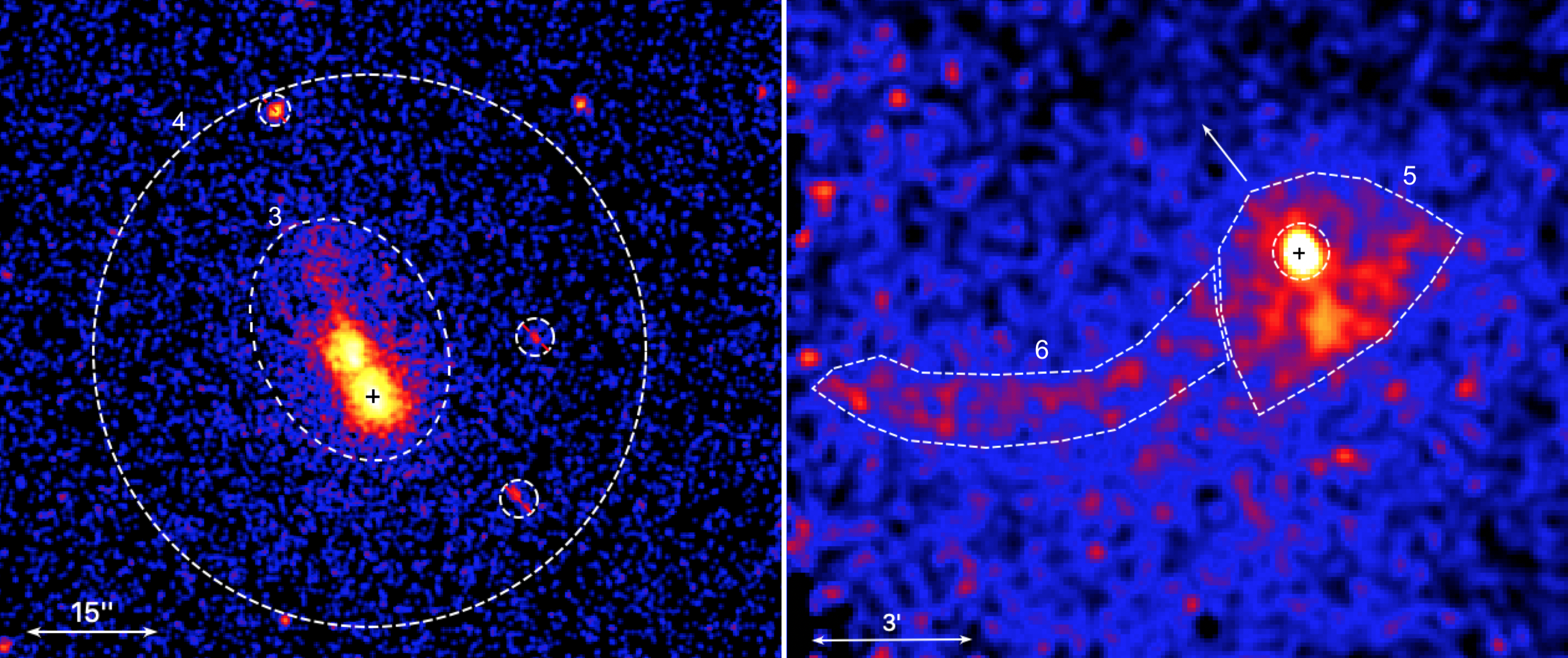}
\caption{{\sl Left:} Merged ACIS image of the CN and its vicinity, binned by a factor of 0.5 and smoothed with a 3-pixel ($r=0\farcs74$) Gaussian kernel.  The following region is shown:  4 -- CN vicinity (the encircled area, excluding the region 3 ellipse and point sources).  {\sl Right:} Merged exposure-map-corrected ACIS image (540 ks) of the J1809 extended nebula, binned by a factor of 10 and smoothed with a 3-pixel ($r=14\farcs9$) Gaussian kernel.  Point sources were removed to enhance the clarity of the diffuse structures.  The arrow represents the presumed direction of pulsar motion (based on the PWN symmetry axis), and the cross marks the pulsar position.  Shown are the following regions:  5 -- EN (the polygon on the right, excluding the region 4 circle);  6 -- outflow (the polygon on the left). Note: the region 6 contour is only a boundary used for spectral extraction; it does not represent the physical boundary of the structure, the structure could extend into region 5 (in 3D).}
\label{fig-med-large}
\end{figure*}

In Figure \ref{fig-med-large} we present the merged exposure-map-corrected images of the CN's vicinity (left panel) and also of the large-scale PWN (right panel).
The deep 536 ks exposure image reveals two interesting features.
The first is that the EN is confined within a dome-like boundary, with its axis coincident with that of the CN.
The EN extends approximately $1\farcm5$ NE of the pulsar, and $2\farcm1$ SW of the pulsar, after which the brightness drops abruptly to background levels.
At its SW boundary (i.e., the widest part) the EN spans 5$'$ across.
The EN's average net surface brightness is $0.214\pm0.007$ counts arcsec$^{-2}$ in the 536 ks exposure.

The second interesting feature is a very elongated and slightly curved outflow extending  eastward from the EN. 
The outflow maintains an average width of roughly $1'$ as it extends at least $7\farcm5$ away from the EN's edge, while curving slightly to the North over its extent.
It is unclear if the outflow's length exceeds $7\farcm5$, or if is just no longer detected beyond that point due to the decreased sensitivity at the considerable off-axis angle and/or extending beyond the detector's field of view.
The outflow's average net surface brightness is less than half that of the EN, at $0.094\pm0.007$ counts arcsec$^{-2}$.

\subsubsection{Spectrum}
All regions of extended emission are well-fit by absorbed PL models (see Table \ref{table-pwn-spectra}).
While evidence of spectral softening is seen between the pulsar vicinity (region 2), the CN (region 3), and the CN vicinity (region 4), no further spectral changes are seen with increasing distance from the pulsar.
The spectrum of the CN's vicinity, the EN (region 5), and the elongated structure (region 6) all exhibit virtually the same spectrum, with $\Gamma_4=1.72\pm0.08$, $\Gamma_5=1.74\pm0.04$, and $\Gamma_6=1.74\pm0.12$.

\section{DISCUSSION}

\subsection{Pulsar Motion and PWN Morphology}

\begin{figure*}
\centering
\includegraphics[width=0.98\hsize,angle=0]{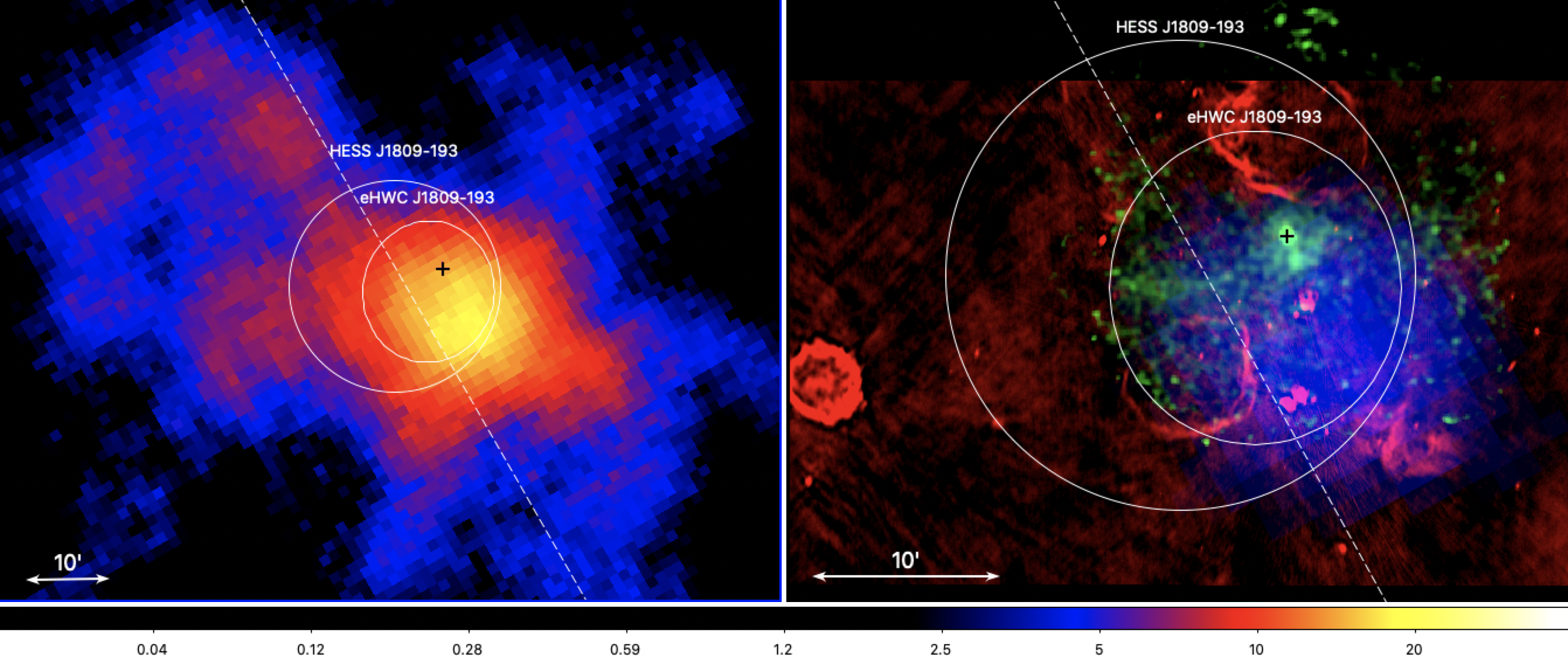}
\caption{\footnotesize {\sl Left:} TeV image from the HESS Galactic Plane Survey \citep{HESS2018} of the HESS J1809--193 field.  The color bar below corresponds to the significance.  {\sl Right:} Multiwavelength image of the J1809 field. Red: radio (JVLA, 1.4 GHz; \citealt{Castelletti2016}), Green: X-ray ({\sl Chandra}-ACIS, 0.5--8 keV), Blue: TeV $\gamma$-rays (the same image from the left panel).  The white ellipses shows the position and positional uncertainties of HESS J1809--193 (which was fit with a 3-point Gaussian; see Table A7 of \citealt{HESS2018}) and the extended HAWC source eHWC J1809--193 (see Table 1 of \citealt{HAWC2019}).  The dashed line marks the Galactic plane ($b=0^\circ$), and the cross marks the position of PSR J1809--1917.}
\label{fig-mw}
\end{figure*}

Although the pulsar's proper motion was only marginally detected, the morphology of EN may be caused by the (transverse) pulsar motion toward the Northeast.
The alignment of the PWN's symmetry axis (i.e., both the CN and EN) with the direction toward the peak of the TeV source HESS J1809--193 (which is located about 7$'$ to the SW\footnote{Note that the peak of HESS J1809 emission is located toward the edge of its 95\% positional uncertainty.}), 
suggests that HESS J1809 is the relic PWN and TeV counterpart of PSR J1809--1917, as was suggested by \citet{Aharonian2007,Kargaltsev2007,HESS2018} and K+18 (see Figure \ref{fig-mw}). 
The center of the extended HAWC source eHWC J1809--193 \citep{HAWC2019,Malone2019} coincides well with both the J1809 PWN and the center of HESS J1809--193, suggesting that eHWC J1809 and HESS J1809 are the same source, and that it is the counterpart to the J1809 PWN.
In Figure \ref{fig-mw} we present a multiwavelength (MW) image of the J1809 field.

Although the dome-like shape of EN resembles the compact heads of SPWNe (see, e.g., \citealt{Kargaltsev2017}) it can hardly be attributed to a bow shock, since the stand-off distance ($r\approx1\farcm5$) is far too large for reasonable values of ISM sound speed or density.
In supersonic pulsars, the bow shock stand-off distance $r_{\rm BS}$ can be crudely estimated by balancing the isotropic pulsar wind pressure\footnote{Although pulsar wind is usually anisotropic in the vicinity of the pulsar, in this case, the rounded leading edge of the EN suggests that the wind flow becomes isotropized at large distances.} with the ram pressure of the ambient medium: $\mu m_{\rm H} n v_{\rm psr}^2 = \dot{E} / (4 \pi c r_{\rm s}^2) $ (where $m_H$ is the mass of the Hydrogen atom, $n$ is the ISM number density, and $\mu=1.3$ is used to convert H density to total density of the ambient medium). With the distance to the apex of the dome $r_{\rm s}=4.4\times10^{18}$ cm ($1\farcm5$ at 3.3 kpc) we get an implausibly low $n\sim 10^{-3} (v_{\rm psr}/100~{\rm km/s})^{-2}$ cm$^{-3}$ (for a typical pulsar speed). Thus, the observed size of the EN cannot be reconciled with supersonic pulsar motion in a normal ambient medium. In the case of transonic motion, the above approximation is no longer applicable, and therefore, an implausibly low ambient medium density is not required.
Given the size of the EN at its estimated distance, the pulsar is likely moving transonically through the ISM (i.e., Mach number $\mathcal{M}\approx 1$).
In the absence of any evidence suggesting that the pulsar resides in the hot interior of a SNR, it is reasonable to assume that the ambient medium (i.e., ISM) sound speeds $c_{\rm ISM}$ are on the order of a few or a few tens of km s$^{-1}$ for the ``cold'' and ``warm'' phases of the ISM (see p.~525 of \citealt{Cox2000}).
A transonic pulsar velocity does not contradict our marginal proper motion detection given its large uncertainties.

\begin{figure*}
\includegraphics[width=0.98\hsize,angle=0]{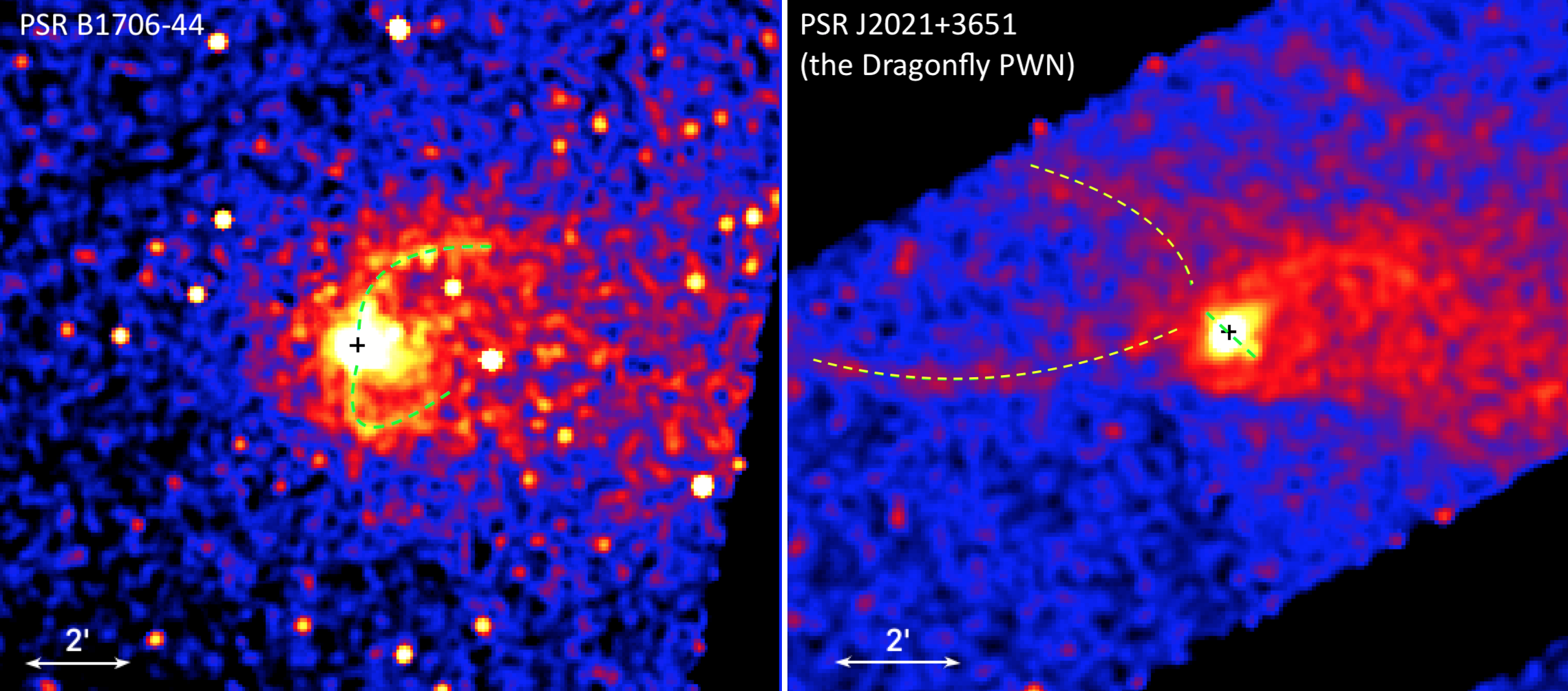}
\caption{{\sl Chandra} ACIS images of the PWNe produced by PSRs B1706--44 (left; 99 ks; ObsID 4608; \citealt{Romani2005}) and J2021+3651, a.k.a.\ the Dragonfly PWN (right; 112 ks; ObsIDs 3091, 7306, and 8502; \citealt{vanEtten2008}).
Although only B1706's motion has been measured (de Vries et al.\ in prep), the large distance between J2021 and its apex indicates that it is also transonic.  Both images were exposure-map-corrected, binned by a factor of 8, and smoothed with a 3-pixel ($r=11\farcs8$) Gaussian kernel. The black crosses mark the pulsar positions, the jets are highlighted by the dashed green lines, and the outflows of the Dragonfly PWN (right panel) are highlighted with dashed yellow lines.}
\label{fig-transonic}
\end{figure*}

The large distance between the pulsar and the northeastern edge of the EN ($r\approx1.4$ pc) is comparable to those of two other known transonic PWNe produced by PSRs B1706--44 ($r\approx1.7$ pc; $d=3$ kpc; \citealt{Romani2005}, de Vries et al.\ in prep) and J2021+3651 ($r\approx1.0$ pc; the Dragonfly PWN; $d\sim3-4$ kpc; \citealt{vanEtten2008}); see Figure \ref{fig-transonic}.
In all three of these PWNe, the pulsar jets are not destroyed by the ram pressure, and remain within the extent of the paraboloid-shaped PWN head (as opposed to tracing the boundaries of bow shock as is seen in some supersonic PWNe; see, e.g., the J1509--5850 PWN, \citealt{Klingler2016a}).

The best-fit value for the proper motion implies that the projected pulsar velocity is $370$ ($\pm170$) km s$^{-1}$ at the DM distance $d=3.3$ kpc, which is at odds with the interpretation of the PWN's morphology (which suggests a much lower transonic pulsar speed).
At such a high velocity, the CN should have been crushed by the ram pressure for any reasonable ISM sound speed (i.e., $\lesssim$ a few tens of km s$^{-1}$).   
Of course, the pulsar velocity could be smaller if the pulsar is closer than its DM distance of 3.3 kpc, but the large $N_{\rm H}$ does not support a significantly smaller distance. 
Using the Parkes radio telescope and a baseline (in time) of 10 years, \citet{Parthasarathy2019} measured a proper motion for J1809 of $\mu_\alpha=-19\pm6$ and $\mu_\delta=50\pm90$ mas yr$^{-1}$ (95\% CL).
Although their measurement appears somewhat discrepant from ours in that they only detect proper motion in the westward direction, it is still consistent within $2\sigma$ of our measurement.
The transverse velocity in R.A.\ computed from their measurement, $v_\alpha=-300\pm100$ km s$^{-1}$, is reasonable with respect to the transverse velocities of the general pulsar population \citep{Verbunt2017}.
K+18 previously reported a velocity of $\mu_\delta=39.6 \pm 8.8$ mas yr$^{-1}$ (using the first 3 epochs of {\sl Chandra} data).
Although also consistent within $2\sigma$ of our measurement, it is possible that they underestimated the systematic uncertainties in the {\sl Chandra} WCS.

\subsection{Variability and the CN}
\begin{figure}
\includegraphics[width=0.98\hsize,angle=0]{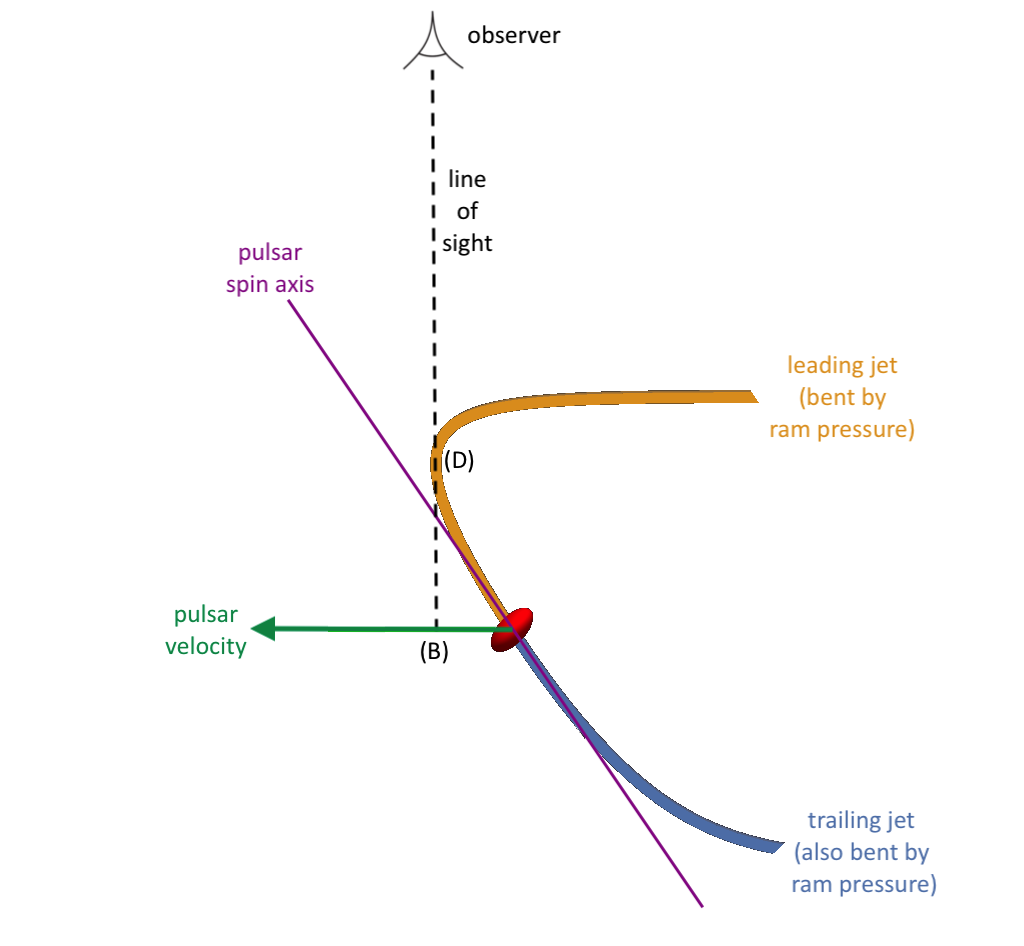}
\caption{Diagram qualitatively illustrating the relative orientations of the line of sight, pulsar velocity, and spin axis.  The maximally Doppler-boosted portion of the leading jet occurs at point (D): the tangent point between our line of sight and the bent leading jet.  Point (B) represents the apparent location of the blob when the Doppler-boosted portion of the jet is projected onto the plane of sky.}
\label{fig-diagram}
\end{figure}

The lack of steady motion of the ``blob'' suggests that the blob is not an actual (isolated) clump of plasma moving away from the pulsar, such as those produced by PSRs J1509--5850 \citep{Klingler2016a} or  B1259--63 \citep{Pavlov2015,Hare2019}.
Instead, it could be a Doppler-boosted region of a bent pulsar jet.
We present a qualitative illustration of the geometry of this scenario in Figure \ref{fig-diagram}.
In this scenario, the leading jet (ahead of the pulsar) is initially launched along the spin axis whose direction is offset from the line of sight.
As the leading jet experiences the ram pressure produced by the pulsar's motion through the ISM, it will get deformed and bent backward.
At some point along the jet's curved path, the jet flow is directed toward the observer. 
When projected onto the sky this part of the jet looks brighter because of Doppler boosting, which could explain the bright clump a few arcseconds NE of the pulsar.
The changing position of the Doppler-boosted region seen in epochs 4--9 and the ``wagging-tail'' motion seen in epochs 1--3 (reported by K+18) can be due to pressure nonuniformities in the region ahead of the pulsar, varying jet flow pressure, kink instabilities within the jet, jet precession, or some combination of the above.
This geometrical interpretation also explains why the CN is extended in the direction ahead of the pulsar, as the trailing jet is initially oriented away from our line of sight and therefore Doppler deboosted (making the CN appear brighter NE of the pulsar).
This picture and interpretation are broadly in agreement with those for the jets of the Vela pulsar \citep{Pavlov2003,Durant2013}.

The blob's spectrum ($\Gamma=1.34\pm0.06$) is marginally softer than the spectrum in the pulsar vicinity ($\Gamma=1.23\pm0.09$), but distinctly harder than the CN's spectrum (region 3; $\Gamma=1.52\pm0.03$).
A similar spectral behavior is seen in the Vela PWN (which is much closer and whose structure is well-resolved; \citealt{Durant2013}).
These support the idea that the blob is a part of the pulsar's jet.

The interpretation of the J1809 PWN and magnetospheric geometry implies that J1809 may be a rather rare instance of a jet-dominated PWN (similar to the PWN of J1811--1925 in SNR G11.2--0.3; \citealt{Roberts2003}). 
Note that both pulsars are $\gamma$-ray quiet, which is expected for the case when the spin and magnetic dipole axes are more or less aligned.  
This may be the reason that some PWNe are jet-dominated (e.g., \citealt{Buhler2016}).  
The J1809 torus, if present, would then be underluminous compared to other PWNe produced by pulsars with similar ages and spin-down luminosities.

\subsection{Misaligned Outflow and Extended Emission}

Extending $\geq$7$'$ from the eastern edge of the EN is a faint collimated structure revealed by the merged image (536 ks).
Although the collimated structure is fainter than the EN ($0.094\pm0.008$ versus $0.214\pm0.007$ net cts arcsec$^{-2}$), both the EN and the structure exhibit identical spectral indices, $\Gamma=1.74$ ($\pm0.05$ and $\pm0.12$, respectively), indicating that they are comprised of electron populations with the same SED. 
The lack of spectral changes could suggest either that there is no noticeable synchrotron cooling occurring as particles travel along the outflow, or that there is a reacceleration mechanism occurring along the outflow preventing any noticeable spectral changes from synchrotron losses (see, e.g., \citealt{Xu2019}).
The outflow's spectrum is also consistent with the spectra exhibited by the outflows produced by the fast-moving pulsars mentioned in Section 1, of which almost all exhibit a photon index in the $1.6-1.7$ range.

This structure could be interpreted as another instance of a ``misaligned outflow'' (a.k.a.\ a ``kinetic jet''; see \citealt{Kargaltsev2017}, \citealt{Reynolds2017}, \citealt{Bykov2017}, \citealt{Barkov2019}, and \citealt{Olmi2019}).
As most of the known misaligned outflows are associated with  supersonic pulsars, one might assume that a high Mach number is required to produce such a structure. 
However, since J1809 is unlikely to have a high Mach number, the existence of a ``kinetic jet'' in this PWN (and also in the transonic Dragonfly PWN; see the right panel of Figure \ref{fig-transonic}) would imply that supersonic velocities are not required to produce such outflows.  
Alternatively, the elongated feature could be analogous to the X-ray filament near the Vela pulsar that is sometimes referred to as Vela X, which is also misaligned with respect to the Vela pulsar's direction of motion, and believed to be formed due to the interaction between the pulsar wind and the reverse shock in the Vela SNR (though Vela X is much larger than this X-ray filament; see \citealt{Abramowski2012}). 
However, J1809 is likely to be older than the Vela pulsar, and no evidence of an SNR associated with J1809 has been found so far \citep{Castelletti2016}.

In the case of J1809 (and also PSR J2021+3651; see Figure \ref{fig-transonic}), the particle leakage can not be the result of large gyroradii \citep{Bandiera2008}, since the gyroradii of X-ray emitting particles here are orders of magnitude smaller than the size of the EN.
For X-ray synchrotron-emitting electrons, the gyroradius $r_g$ can be estimated as $r_g \sim 10^{14}(E/1\ {\rm keV})^{1/2} (B/20\ {\rm \mu G})^{-3/2}$ cm.
For an 8-keV-emitting electron in a 5 ${\rm \mu G}$ magnetic field\footnote{The magnetic field strength in the ISM is typically on the order of a few ${\rm \mu G}$; so the magnetic field strengths inside PWN would have to be greater due to the presence of the magnetized wind.}, $r_g \approx 2.3\times10^{15}$ cm.
At the distance of J1809 ($d\sim3.3$ kpc), this corresponds to $\approx0.05''$.
Thus, this case can be interpreted as observational evidence that PWN and ISM magnetic fields can reconnect in the nebulae of transonic pulsars as well as supersonic pulsars.
If this structure is produced by PWN/ISM magnetic field reconnection, the point of reconnection is not necessarily at the interface between regions 5 and 6 as shown in Figure \ref{fig-med-large}, since that is a 2D projection of a 3D structure, but could be at a point ``inside'' (in projection) region 5.
However, in the MHD simulations which demonstrated this reconnection scenario \citep{Barkov2019,Olmi2019}, the pulsars had Mach numbers $\mathcal{M}\gg1$, so further simulations for the transonic case are needed.

\section{Point Sources in the J1809 Field}

In order to elucidate the relationship between PSR J1809--1918 and HESS/eHWC J1809--193, we utilized the half-megasecond {\sl Chandra} exposure of the J1809 field to search for and identify any other sources capable of contributing to the extended TeV emission of HESS/eHWC J1809-193.
The most likely location of the central engine of the TeV emission is the overlap between the HESS and HAWC localizations, which (in this case) is the HAWC 95\% error region (fully covered by the {\sl Chandra} observations).
It is also possible that the TeV emission in the J1809 field has multiple contributing sources (in addition to PSR J1809). 
Therefore, we examine the other X-ray sources in the field in order to determine if any of them could also be contributing to the TeV emission.

Another reason to explore the field sources is that this is a rich galactic field, and therefore, there is chance of finding interesting and/or unusual sources  (even if they are unrelated to the J1809 PWN or TeV emission).

\subsection{Source Detection \& Classification}
\label{srcdet}
We analyzed  15 {\sl Chandra} observations covering the J1809 field and listed in Table \ref{tbl-obs}. We excluded ObsIDs 20332 and 20967 because the astrometric correction did not converge to the required precision.  
The 2$\sigma$ statistical positional uncertainties (PU) were calculated using the empirical relation between PU, off-axis angle, and source counts from \citet{Kim2007}.
The 2-$\sigma$ astrometric uncertainty was then added in quadrature to the statistical PU to calculate   the overall $2\sigma$ PU. 

We ran the CIAO tool {\tt srcflux} on each observation, using the source coordinates found by {\tt wavdetect} from the astrometrically corrected images.
The sources detected with a $\ge 6\sigma$ significance were then selected and cross-matched (within $2\farcs5$) between observations. After excluding sources located within $1\farcm0$ of the pulsar/PWN region (to avoid PWN emission clumps being interpreted as point sources), 90 unique sources remained, with many of them detected (at $\ge 6\sigma$) in multiple observations. Another cut was then applied to keep only those sources that are detected at a $>10\sigma$ confidence level in at {\em least one observation}.
The more demanding 10$\sigma$ cut is needed to ensure that the source is detected significantly enough so that the fluxes in the individual bands and corresponding hardness ratios can be reliably calculated in at least on observation. 
On the other hand, the $>6\sigma$ detections in other observations are sufficient for comparisons of broadband fluxes to probe each source's variability.
After the above cuts were applies we were left with 35 sources (with 199 total detections across all 15 observations).

We then found weighted averages of the positions and fluxes of each source, using the inverse square uncertainties as weights. To identify variable sources, we fitted a constant flux model to the broad band (0.7-7.0 keV) fluxes measured from multiple observations and calculated the minimum reduced chi-square, $\chi_\nu^2$.
The X-ray fluxes, $F_s$, $F_m$, $F_h$, and their uncertainties are measured in three energy bands ($s$ = 0.7 - 1.2, $m$ = 1.2 - 2, $h$ = 2.0 - 7 keV) for each source. 
The broad band flux $F_b$ ($b$ = 0.7 - 7 keV) is calculated by co-adding the fluxes at soft, medium and hard bands
together. 
The uncertainty of $F_b$ is calculated from the flux uncertainties in each of the three bands added in quadrature\footnote{Note that CIAO's {\tt srcflux} reports ``NaN'' for the model-independent flux in the case of very low or zero counts (see \url{https://cxc.harvard.edu/ciao/ahelp/srcflux.html}). 
We treated these NaN values as 0 while calculating $F_b$. }.
These energy bands are chosen because they overlap with the energy bands used in the 3XMM-DR8 catalog\footnote{See \url{http://xmmssc.irap.omp.eu/Catalogue/3XMM-DR8/3XMM-DR8_Catalogue_User_Guide.html}}, which we use to construct our training dataset (see Appendix). 
Due to the increasing buildup of a contaminating layer on the ACIS detector (see e.g., \citealt{Plucinsky2018}), and uncertainties in the ACIS quantum efficiency contamination model near the O-K edge at 0.535 keV and F--K edge at 0.688 keV\footnote{See \url{ https://cxc.cfa.harvard.edu/ciao/why/acisqecontamN0010.html}}, we chose to use $0.7-1.2$ keV for the soft energy band.  
From these fluxes we calculated two hardness ratios HR2$=(F_m-F_s)/(F_m+F_s)$ and HR4$=(F_h-F_m)/(F_h+F_m)$. 
The plot of  hardness ratios of all 35 sources is shown in Figure~\ref{hr_diagram} together with the sources from the training dataset (see Appendix).
These and other X-ray parameters for each source are also listed in Table ~\ref{tabsources}.

The multi-wavelength (MW) photometry was taken from the {\sl Gaia} DR2 catalog \citep{Gaia2018}, the Two Micron All Sky Survey (2MASS; \citealt{Skrutskie2006}), and the Galactic Legacy Infrared Midplane Survey Extraordinaire (GLIMPSE; \citealt{Benjamin2003}). 
These catalogs were cross-matched with each source's averaged X-ray position and were considered to be a match if the MW counterpart was located within the 2$\sigma$ X-ray PU. 
We found that 18 out of 35 X-ray sources had MW counterparts. 
The chance coincidence probabilities for each catalog were calculated by finding the average field source density, $\rho$, and calculating the probability, $P=1-\exp{(-\rho\pi r^2)}$, of having one or more sources within a randomly placed, $r=1''$ circle, which is the average 2$\sigma$ PU of all of the sources. 
The source densities from the corresponding catalogs are $\rho_{\rm Gaia}=0.0147$ arcsec$^{-2}$, $\rho_{\rm 2MASS}=0.0105$ arcsec$^{-2}$, and $\rho_{\rm GLIMPSE}=0.0245$ arcsec$^{-2}$, leading to chance coincidence probabilities of $4.5\%$, $3.2\%$, and $7.4\%$, respectively. 
The GLIMPSE catalog has the largest chance coincidence probability, which implies that $\sim3$ GLIMPSE sources may be spurious cross-matches.
None of the X-ray sources had more than one {\sl Gaia} or 2MASS counterpart; however, one X-ray source (Source 26) had two potential GLIMPSE counterparts, so the nearest match was taken as the counterpart.

The magnitudes and colors for the MW counterparts, together with the X-ray properties (fluxes in two bands and two hardness ratios), add up to 20 features, which were used to classify these sources with a machine learning (ML) approach (see Appendix; \citealt{Hare2016,Hare2017}. 
The summary of the MW properties of the sources and their classifications are given in Table ~\ref{tabsources}, with 13 sources (marked in bold) having classification confidence $>70\%$ (see Appendix). 
At this level of confidence, we do not identify any Galactic objects capable of contributing to the extended TeV emission of HESS/eHWC J1809--193 (e.g., pulsars or HMXBs). 
It is worth noting that there are several confidently identified AGNs that could, in principle, produce some TeV emission. 
However, in X-rays they are significantly fainter than those AGNs that are known to be TeV sources.

\begin{figure*}
\centering
\includegraphics[width=0.98\hsize,angle=0]{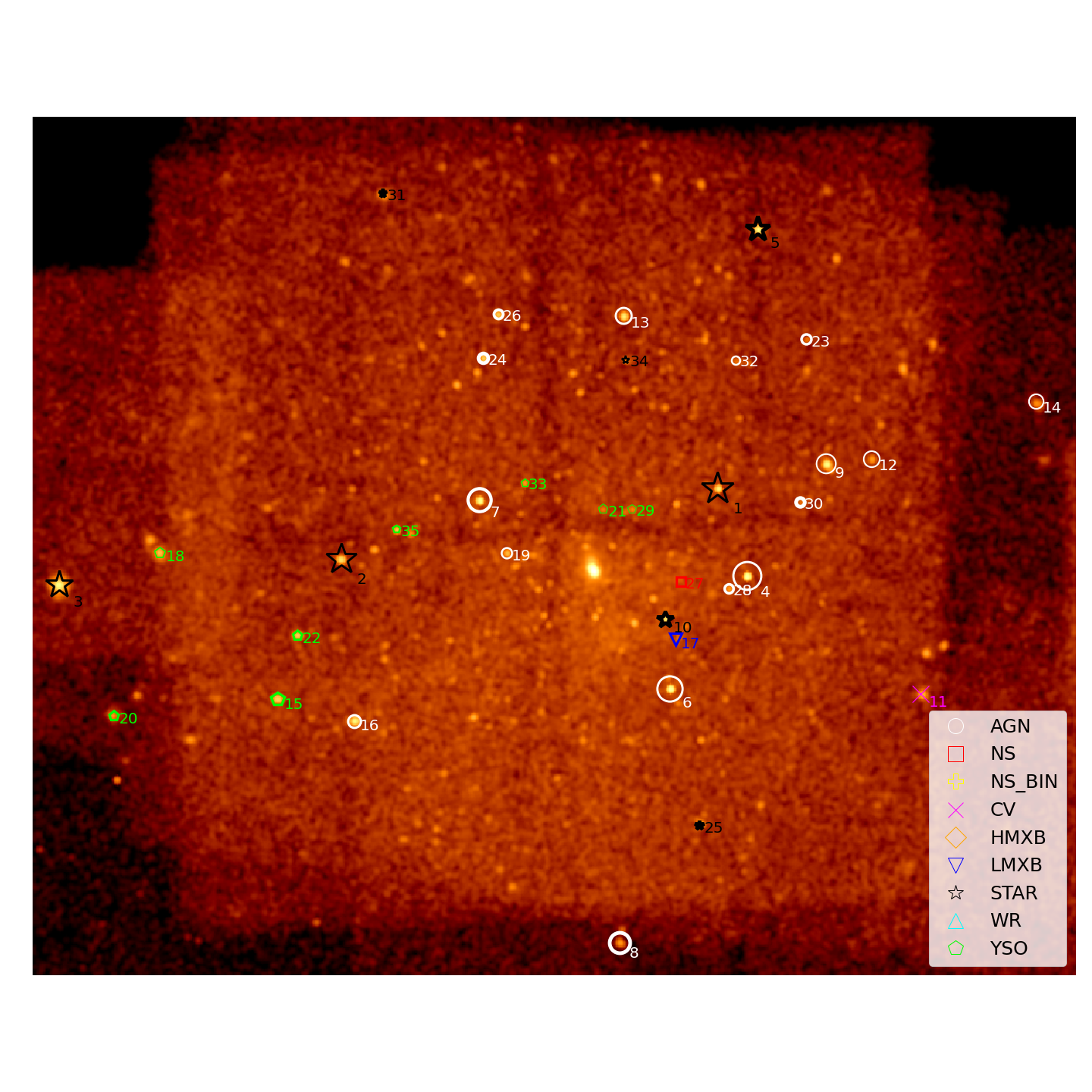}
\caption{\footnotesize X-ray image of the J1809 field and the 35 classified  sources  (see text and Table~\ref{tbl-obs}). The sizes of the symbols are proportional to the source detection significance, while the line thickness is proportional to the classification confidence. The symbols mark the class of the X-ray source based on the ML classification (see the legend and Table~\ref{tabsources}: active galactic nucleus, neutron star, neutron star binary, cataclysmic variable, high mass X-ray binary, low mass X-ray binary, star, Wolf-Rayet star, young stellar object).}
\label{source_image}
\vspace{-0.0cm}
\end{figure*}

The X-ray image of the J1809 field, constructed by stacking all 15 observations together, is shown in Fig.~\ref{source_image}. 
All 35 sources with a detection significance $>10$ are shown by colored circles.
The size of each circle is proportional to each source's detection significance, while the circle's thickness is proportional to the source's classification confidence.

Since our ML classification currently does not take into account temporal properties, and TeV HMXBs could be variable in X-rays, we examined the two brightest, significantly variable ($>5\sigma$) and transient (i.e., only significantly detected in a single observation) sources.
For sources with an optical/NIR counterpart, we use MW photometry to fit the SED of each source to obtain a spectral type using the VO SED Analyzer (VOSA; \citealt{Bayo2008}). 
The spectral type is then determined by using a chi-square fit to the NextGen theoretical model spectra \citep{Hauschildt1999}.
We also include the {\sl Gaia} derived distance estimates in the SED fits when available \citep{Bailer-Jones2018}.

\begin{figure*}
\centering
\includegraphics[width=0.9\hsize,angle=0]{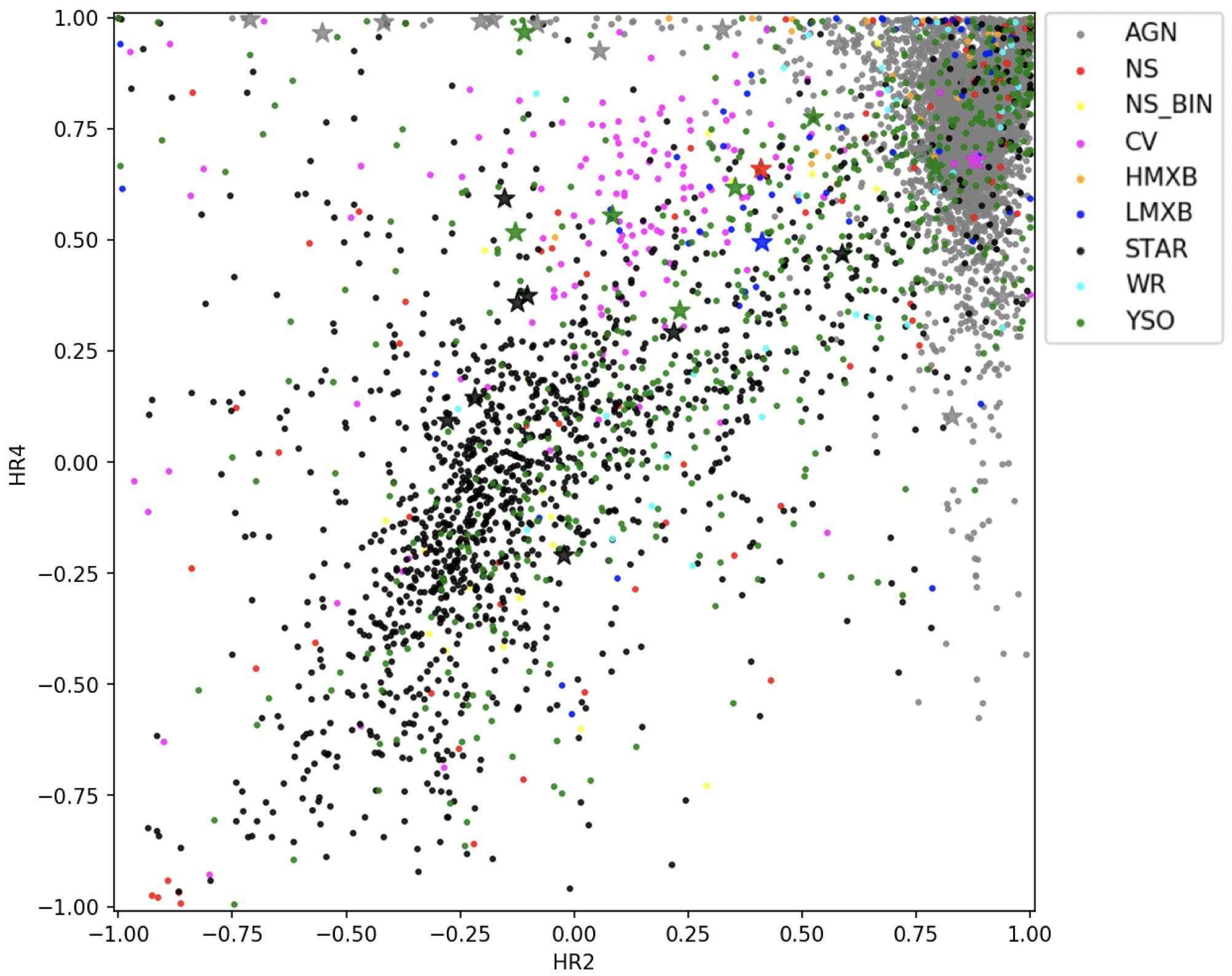}
\caption{\footnotesize The hardness ratio diagram (HR2 versus HR4) with the training dataset (circles) and classified sources  from the J1809 field (stars). The color shows the source class according to the legend.}
\label{hr_diagram}
\vspace{-0.0cm}
\end{figure*}

\subsection{Variable Sources}

Sources 1 and 2 both show strong X-ray variability across the 17 observations, but neither source is particularly variable within the duration of a single observation. 
The {\sl Gaia} distances to these sources are 3 kpc and 2 kpc, respectively, but we caution that both are impacted by large astrometric noise (possibly due to the sources being binaries), which may cause these distance estimates to be unreliable \citep{Bailer-Jones2018}. 
At these distances, both sources have observed peak X-ray luminosities of  $\sim8\times10^{31}$ erg s$^{-1}$, and ``quiescent'' luminosities of $8\times10^{30}$ erg s$^{-1}$ and $2\times10^{30}$ erg s$^{-1}$, respectively.
Both sources also have similar parameters derived from fits to their NIR-optical SEDs (i.e., $T_{\rm eff}\approx4500$ K and $\log g=3.5$), implying cool giants\footnote{We note that the fitted spectral models assume solitary stars, and may be inaccurate if the sources are binaries.}.  
Therefore, neither of these two sources are likely to be HMXBs, nor are they likely to be responsible for the TeV emission. 
The hardness ratios indicate soft spectra with only modest absorption (in agreement with NIR-optical SED shape) suggesting that these are not quiescent LMXBs, which tend to have hard X-ray spectra.  
Since the X-ray luminosities are somewhat larger than stellar (for these type of stars), we believe that both sources could be active binaries (ABs), similar to, e.g., W Uma, RS CVn \citep{vandenBerg2013}, which are known to exhibit flares with luminosities up to $10^{34}$ erg s$^{-1}$ (e.g., \citealt{Tsuboi2016}).

\subsection{Transient Sources}

The two brightest short transient sources detected in the {\sl CXO} observations are Sources 12 and 17. 
Source 12 only shows up in ObsID 21098 and does not exhibit strong variability  throughout the duration of the observation. 
No optical/NIR counterpart was found for this source in the catalogs used for ML classification; however, a faint NIR source is detected in the UKIDSS Galactic plane survey \citep{Lucas2008}. 
The source is offset by 0\farcs4, and has magnitudes $J=18.4$, $H=16.4$, and $K=15.5$, suggesting that it is strongly absorbed. 
At a distance of 4--10 kpc, the source would have an observed X-ray luminosity of $10^{32}-10^{33}$ erg s$^{-1}$, which is still consistent with being a flaring AB type source, or, if the source is more nearby, with a flaring M-dwarf star (see e.g., \citealt{Stelzer2013}, \citealt{Pye2015}). 
It could also be a flaring AGN, given the hard spectrum.  
Source 17 is only detected in ObsID 20330, and shows a single flare, which decays over $\sim7$ ks to a persistent low flux level. 
The source has no MW counterpart\footnote{We note there are four VPHAS+ NIR sources \citep{Drew2016} located $0\farcs9-1\farcs25$ offset from Source 17, so it may be that there are blended sources making it difficult to detect a possibly faint counterpart to Source 17.}, and there are too few X-ray photons (38$\pm6$ in total with a mean energy of $\sim2.4$  keV) to constrain the spectrum and nature of this relatively hard source without knowledge of its distance. 

Lastly, Source 27 is a short transient only detected in ObsID 21126. 
This source shows flaring behavior, with the longest duration flare lasting $\sim 2$ ks. 
The total number of photons is small (32$\pm6$ in total) precluding any meaningful spectral analysis. 
There is one potential counterpart in the UKIDSS Galactic plane survey that lies just at the edge ($\sim0\farcs65$) of the source's PU and has magnitudes $J=18.5$, $H=17.6$, and $K=17.1$, and is classified as a galaxy in the UKIDDS catalog. 
If this source is the true counterpart, then Source 27 could be an AGN.
We note that the ML-based classification of this source, which is not confident (i.e., $<70\%$), as an isolated NS (e.g., pulsar) is unlikely since the source is strongly variable. 
However, the current training dataset does not take into account temporal information.
To conclude, none of the transient sources appear to be a good HMXB candidate (which could contribute to the TeV flux) and, although AGN classification is not excluded for some of these sources, the low X-ray flux would imply low TeV flux in this scenario. 
Therefore, PSR J1809--1917 remains the most likely counterpart to HESS/eHWC J1809--193.

\begin{deluxetable*}{lcccccccccccc}
\tablecolumns{13}
\tablecaption{{\sl CXO} Sources Detected in the J1809 Field}
\tablehead{
\colhead{$\#$\tablenotemark{a}} &  \colhead{R.A.} & \colhead{Dec.} & \colhead{PU\tablenotemark{b}} & \colhead{SNR\tablenotemark{c}} & \colhead{$\chi^2/\nu$\ \tablenotemark{d}} & \colhead{$F_b$\tablenotemark{e}} &  \colhead{$HR2$\tablenotemark{f}} &  \colhead{$HR4$\tablenotemark{f}} &   \colhead{Gmag\tablenotemark{g}} &   \colhead{Jmag\tablenotemark{h}} & \colhead{5.8mag\tablenotemark{i}} &  \colhead{Class(\%)\tablenotemark{j}} }
\startdata
1 & 272.38184 & -19.26356 &  0.57 &        54.03 &       58.9/13 &    1.35$\pm$0.32 &   $0.59^{+0.21}_{-0.25}$ & $0.47^{+0.17}_{-0.21}$ &  16.31 &  12.51 &  10.71 &  STAR(?) \\
2 & 272.52743 & -19.28941 &  0.75 &        52.40 &       57.2/6 &    0.85$\pm$0.22 &   $0.22^{+0.32}_{-0.32}$ & $0.29^{+0.24}_{-0.29}$ &  17.89 &  14.21 &  10.12 &  STAR(?) \\
3 & 272.63647 & -19.29869 &  1.67 &        46.26 &        ... &   24.39$\pm$1.58 &  -$0.10^{+0.08}_{-0.07}$ & $0.37^{+0.06}_{-0.06}$ &  12.81 &  10.87 &   9.55 &  STAR(?) \\
4 & 272.37050 & -19.29553 &  0.54 &        44.43 &       12.1/10 &    7.02$\pm$1.20 &  -$0.21^{+0.54}_{-0.48}$ & $0.99^{+0.00}_{-0.01}$ &    ... &    ... &    ... &   AGN(?) \\
5 & 272.36630 & -19.16929 &  0.77 &        40.29 &       51.4/11&    2.56$\pm$0.57 &  -$0.22^{+0.24}_{-0.20}$ & $0.15^{+0.23}_{-0.30}$ &  14.25 &  12.63 &  11.68 &  \textbf{STAR(98)} \\
6 & 272.40015 & -19.33660 &  0.54 &        39.74 &       21.7/13 &    4.93$\pm$0.94 &   $0.58^{+0.25}_{-0.43}$ & $0.94^{+0.03}_{-0.04}$ &    ... &    ... &    ... &   AGN(?) \\
7 & 272.47389 & -19.26788 &  0.62 &        35.83 &       20.8/11 &    3.08$\pm$0.71 &   $0.33^{+0.39}_{-0.63}$ & $0.97^{+0.01}_{-0.02}$ &    ... &    ... &    ... &   \textbf{AGN(86)} \\
8 & 272.41968 & -19.42886 &  0.81 &        31.26 &       10.8/4 &    9.62$\pm$1.40 &   $0.86^{+0.08}_{-0.12}$ & $0.86^{+0.04}_{-0.04}$ &    ... &    ... &    ... &   \textbf{AGN(97)} \\
9 & 272.33989 & -19.25460 &  0.71 &        29.29 &       7.1/9 &    5.60$\pm$1.13 &  -$0.71^{+0.47}_{-0.19}$ &  $1.00^{+0.00}_{-0.01}$ &    ... &    ... &    ... &   AGN(?) \\
10 & 272.40211 & -19.31132 &  0.59 &        25.70 &       18.5/7 &    0.63$\pm$0.20 &  -$0.02^{+0.33}_{-0.30}$ &  -$0.21^{+0.38}_{-0.37}$ &  12.24 &  11.37 &  10.95 &  \textbf{STAR(96)} \\
11 & 272.30315 & -19.33844 &  0.87 &        24.68 &       19.7/4 &    3.54$\pm$0.77 &   $0.88^{+0.07}_{-0.10}$ & $0.68^{+0.11}_{-0.13}$ &  19.90 &    ... &    ... &    CV(?) \\
12$^*$ & 272.32227 & -19.25281 &  1.07 &        23.42 &        ... &    6.88$\pm$1.25 &    ... &  $0.95^{+0.03}_{-0.04}$ &    ... &    ... &    ... &   AGN(?) \\
13 & 272.41819 & -19.20079 &  0.76 &        23.40 &       9.4/7 &    4.17$\pm$0.93 &  -$0.18^{+0.57}_{-0.55}$ & $1.00^{+0.00}_{-0.01}$ &    ... &    ... &    ... &   AGN(?) \\
14 & 272.25864 & -19.23210 &  1.47 &        20.96 &        ... &   16.67$\pm$2.06 &  -$0.08^{+0.52}_{-0.50}$ & $0.98^{+0.01}_{-0.01}$ &    ... &    ... &    ... &   AGN(?) \\
15 & 272.55197 & -19.34022 &  1.45 &        19.68 &       17.0/4 &    2.71$\pm$0.52 &   $0.79^{+0.12}_{-0.21}$ & $0.66^{+0.11}_{-0.12}$ &  19.35 &  14.87 &    ... &   \textbf{YSO(73)} \\
16 & 272.52225 & -19.34824 &  0.99 &        17.75 &       38.5/13 &    4.11$\pm$0.80 &  -$0.42^{+0.55}_{-0.37}$ & $0.99^{+0.01}_{-0.01}$ &    ... &    ... &    ... &   AGN(?) \\
17$^*$ & 272.39803 & -19.31851 &  0.73 &        16.23 &        ... &    1.82$\pm$0.57 &  $0.41^{+0.31}_{-0.39}$ & $0.50^{+0.21}_{-0.27}$ &    ... &    ... &    ... &  LMXB(?) \\
18 & 272.59734 & -19.28699 &  1.39 &        14.79 &        ... &    8.15$\pm$1.20 &    ... &  $0.95^{+0.02}_{-0.03}$ &    ... &    ... &    ... &   YSO(?) \\
19 & 272.46344 & -19.28710 &  0.72 &        13.78 &       0.38/4 &    1.47$\pm$0.52 &  -$0.55^{+0.50}_{-0.30}$ & $0.97^{+0.02}_{-0.05}$ &    ... &    ... &    ... &   AGN(?) \\
20 & 272.61540 & -19.34647 &  1.95 &        13.42 &        ... &    5.52$\pm$0.94 &   $0.35^{+0.17}_{-0.17}$ &  $0.62^{+0.09}_{-0.11}$ &  15.11 &  12.58 &  11.28 &   YSO(?) \\
21 & 272.42610 & -19.27109 &  0.65 &        13.38 &       3.3/1 &    0.32$\pm$0.16 &   $0.08^{+0.43}_{-0.45}$ &  $0.56^{+0.24}_{-0.39}$ &  19.13 &    ... &    ... &   YSO(?) \\
22 & 272.54439 & -19.31711 &  1.41 &        12.92 &       3.5/4 &    2.21$\pm$0.60 &   $0.52^{+0.28}_{-0.50}$ &  $0.78^{+0.11}_{-0.14}$ &  19.68 &  14.20 &  11.20 &   YSO(?) \\
23 & 272.34743 & -19.20938 &  1.15 &        12.67 &        ... &    0.75$\pm$0.23 &   $0.83^{+0.11}_{-0.22}$ & $0.10^{+0.26}_{-0.34}$ &    ... &    ... &    ... &   \textbf{AGN(74)} \\
24 & 272.47246 & -19.21609 &  1.01 &        12.34 &       4.1/7 &    1.11$\pm$0.38 &   $0.76^{+0.15}_{-0.33}$ &  $0.70^{+0.14}_{-0.19}$ &    ... &    ... &    ... &   \textbf{AGN(97)} \\
25 & 272.38873 & -19.38628 &  1.34 &        11.98 &       3.2/1 &    1.72$\pm$0.45 &  -$0.15^{+0.28}_{-0.25}$ & $0.59^{+0.15}_{-0.21}$ &  14.45 &  12.98 &  12.29 &  \textbf{STAR(95)} \\
26 & 272.46664 & -19.20017 &  1.26 &        11.43 &       5.4/3 &    2.19$\pm$0.69 &    ... &   $0.98^{+0.01}_{-0.03}$ &    ... &    ... &    ... &   \textbf{AGN(79)} \\
27$^*$ & 272.39593 & -19.29765 &  0.65 &        11.31 &        ... &    1.33$\pm$0.53 &   $0.41^{+0.32}_{-0.48}$ & $0.66^{+0.18}_{-0.27}$ &    ... &    ... &    ... &  NS(?) \\
28 & 272.37751 & -19.30023 &  0.78 &        10.99 &        ... &    1.04$\pm$0.33 &    ... &  $0.95^{+0.03}_{-0.07}$ &    ... &    ... &    ... &   \textbf{AGN(76)} \\
29 & 272.41503 & -19.27108 &  0.61 &        10.81 &        ... &    0.37$\pm$0.21 &   $0.23^{+0.37}_{-0.41}$ & $0.34^{+0.33}_{-0.49}$ &  18.19 &    ... &    ... &   YSO(?) \\
30 & 272.34974 & -19.26876 &  0.97 &        10.61 &        ... &    0.57$\pm$0.22 &   $0.94^{+0.04}_{-0.15}$ & $0.37^{+0.25}_{-0.34}$ &    ... &    ... &    ... &   \textbf{AGN(99)} \\
31 & 272.51118 & -19.15637 &  2.10 &        10.48 &       9.2/3 &    1.48$\pm$0.32 &  -$0.28^{+0.20}_{-0.19}$ & $0.09^{+0.26}_{-0.34}$ &  15.45 &  13.64 &    ... &  \textbf{STAR(90)} \\
32 & 272.37483 & -19.21707 &  1.35 &        10.12 &        ... &    1.83$\pm$0.55 &   $0.05^{+0.47}_{-0.51}$ & $0.93^{+0.05}_{-0.08}$ &    ... &    ... &    ... &   AGN(?) \\
33 & 272.45647 & -19.26167 &  0.82 &        10.11 &       3.3/1 &    0.30$\pm$0.15 &  -$0.13^{+0.46}_{-0.39}$ & $0.52^{+0.26}_{-0.42}$ &  18.19 &    ... &    ... &   YSO(?) \\
34 & 272.41733 & -19.21707 &  1.21 &        10.09 &        ... &    1.10$\pm$0.53 &  -$0.12^{+0.33}_{-0.33}$ & $0.36^{+0.33}_{-0.48}$ &  17.43 &  14.05 &  10.69 &  STAR(?) \\
35 & 272.50596 & -19.27867 &  0.98 &        10.00 &       4.2/8 &    1.78$\pm$0.55 &  -$0.11^{+0.56}_{-0.53}$ &  $0.97^{+0.02}_{-0.05}$ &    ... &    ... &  12.00 &   YSO(?) 
\enddata
\label{tabsources}
\tablenotetext{ a}{\#  -- Source number (sources are ordered by their significance).}
\tablenotetext{ b}{PU  -- Maximum 2-$\sigma$ positional uncertainties in unit of arcseconds among the observations. }
\tablenotetext{ c}{SNR -- Maximum signal-to-noise ratio among the observations. }
\tablenotetext{ d}{$\chi^2/\nu$ --  Chi-squared  and number of degrees of freedom ($\nu=$number of observations $-1$) for a constant model fitted to the broad band fluxes from all observations. Chi-squared is not calculated for the sources detected only once (transients) or observed only once.  }
\tablenotetext{ e}{$F_b$ -- Averaged broad band flux in the $0.7-7$~keV range in units of $10^{-14}$ erg~s$^{-1}$~cm$^{-2}$ by weighting with the inverse of \\ squares of their 90\% confidence uncertainties.}
\tablenotetext{ f}{HR2 and HR4 -- hardness ratios defined in Section ~\ref{srcdet}}
\tablenotetext{ g}{Gmag -- G band magnitude taken from the GAIA DR2 catalog.}
\tablenotetext{ h}{Jmag -- J band magnitude taken from the 2MASS catalog.}
\tablenotetext{ i}{5.8mag -- 5.8 $\mu$ band magnitude taken from the GLIMPSE catalog.}
\tablenotetext{ j}{Class -- Predicted classification of each source obtained from the automated classification pipeline with their classification confidence (shown in parentheses). Sources with ``?'' have a classification confidence $<70\%$.}
\tablenotetext{*}{Transient sources detected in only one observation. }
\end{deluxetable*}

\section{SUMMARY}
We have presented an analysis of a {\sl Chandra} monitoring campaign of PSR J1809--1917 and its variable pulsar wind nebula.

The pulsar's spectrum is well fit by a power-law + blackbody model, with $\Gamma=1.28\pm0.15$ (possibly due to nebular emission), $kT=0.19\pm0.03$ keV (corresponding to $T=2.2\pm0.4$ MK), and an emitting BB radius of $r=530_{-200}^{+430}$ m (at 3.3 kpc).
The pulsar exhibits a hint of northward motion, $\mu_\delta = 24\pm11$ mas yr$^{-1}$, but only at a 95\% confidence level.
At the pulsar's DM distance, this corresponds to $370\pm170$ km s$^{-1}$.

The arcsecond-scale compact nebula is approximately symmetric about a northeast-southwest axis, and is more elongated toward to the northeast.
The elongation appears to be dominated by emission from a bright ``blob''. 
Between the epochs of the monitoring campaign ($\approx$6--8 week intervals), the blob changes position, size, and brightness, but no steady (linear) motion is seen and the spectrum appears to be constant in time ($\Gamma=1.34\pm0.06$) 
We suggest that the blob is the Doppler-boosted region of a pulsar jet bent toward the observer due to the ram pressure caused by the pulsar's northeastern motion.
The CN appearing shorter toward the southwest could be due to the Doppler-deboosting of the southwest (counter) jet.

The arcminute-scale extended nebula ($\Gamma=1.74\pm0.05$) apparently exhibits a dome-like morphology, with the apex of the dome located $\sim$1.5$'$ NE of the pulsar
The EN's axis of symmetry is the same as that of the CN.  
It is aligned with the direction to the peak of the TeV source HESS J1809--193, located $\approx$7$'$ ($\approx$7 pc) to the southwest.
The entire PWN is located within the 95\% positional uncertainties of the centroids of both HESS J1809--193 and eHWC J1809--193, suggesting that both are the same source, and that HESS/eHWC J1809--193 is the TeV counterpart to the J1809 PWN.
The EN's appearance suggests it is shaped by the pulsar's NE motion, and that the pulsar is moving at a transonic speed.

The deep 536 ks ACIS image reveals a faint narrow structure ($\Gamma=1.74\pm0.12$) extending beyond $7'$ east from the apparent edge of the EN.
The structure could be another instance of a ``misaligned outflow'' or ``kinetic jet'', which have been seen in the PWNe of a handful of supersonic pulsars.
If this is the case, the discovery of such a structure in the J1809 PWN suggests that large Mach numbers are not required to produce these particle leakage phenomena.

We have performed an analysis of the 35 brightest X-ray field sources imaged in these {\sl CXO} observations and found no other plausible sources of TeV emission, thus strengthening the connection between PSR J1809--1917 and  HESS J1809--193  (eHWC J1809--193).
The monitoring observations of this field revealed a number of variable sources (including a few transient sources that varied on timescales of a few ks), which are likely due to interactions in active binaries or flares on single stars (at least for the sources with optical/IR counterparts) given the low fluxes/luminosities involved.

\facility{{\sl CXO}} 
\software{CIAO (v4.11; \citealt{Fruscione2006}), Wavdetect \citep{Freeman2002}, XSPEC (v12.10.1f, \citealt{Arnaud1996})}

\acknowledgements
The authors wish to thank the anonymous referee whose helpful suggestions improved the quality and clarity of the paper. 
The work of OK and GP was supported by the National Aeronautics and Space Administration (NASA) under grant number 80NSSC19K0576 issued through the  NNH18ZDA001N  Astrophysics Data Analysis Program (ADAP).  
Support for this work was also provided by the National Aeronautics and Space Administration through Chandra Award Number GO8-19059 issued by the Chandra X-ray Observatory Center, which is operated by the Smithsonian Astrophysical Observatory for and on behalf of the National Aeronautics Space Administration under contract NAS8-03060.   
JH acknowledges support from an appointment to the NASA Postdoctoral Program at the Goddard Space Flight Center, administered by the USRA through a contract with NASA.

\appendix
\label{sec:method}

In this paper, we use our automated multi-wavelength machine-learning classification pipeline (MUWCLASS; \citealt{Hare2016,Hare2017}), which relies on a Random Forest (RF; \citealt{Breiman2001}) classifier and is implemented in  python \citep{Pedregosa2012}.  
For completeness, we briefly describe the pipeline and the recent modifications below.

We have expanded our training dataset (TD) to $\sim 10,600$ literature-verified sources from the nine predefined classes by cross-matching them with the 3XMM-DR8 catalog sources using a cross-matching radius of $2''$.  
The source types used to construct the training dataset taken from the following catalogs:
\begin{itemize}
\item AGNs were taken from Veron Catalog of Quasars $\&$ AGN 13th Edition \citep{2010A&A...518A..10V}
\item CVs were taken from the Cataclysmic Variables Catalog 2006 Edition \citep{2001PASP..113..764D}
\item LMXBs were taken from Low Mass X-ray Binary Catalog \citep{2007A&A...469..807L}
\item HMXBs were taken from the Catalog of High--Mass X-ray Binaries in the Galaxy 4th Edition \citep{2006A&A...455.1165L}
\item Main sequence stars were taken from the General Catalog of Variable Stars \citep{2009yCat....102025S}
\item NS and binary non-accreting NS were taken from the ATNF Pulsar Catalog \citep{Manchester2005}
\item WR were taken from the VIIth Catalog of Galactic Wolf-Rayet Stars \citep{2001NewAR..45..135V}
\item YSOs were taken from A Pan-Carina YSO catalog \citep{2011ApJS..194...14P}
\end{itemize}
Then the TD sources are cross-matched with the same MW catalogs as those used in \citet{Hare2016} to extract MW parameters within $2''$, except that we now use the {\sl Gaia} DR2 catalog for optical features\footnote{specifically, {\sl Gaia}'s  $G$, $B$, and $R$ band magnitudes.} (instead of the USNO-B catalog) and the GLIMPSE survey catalog \citet{Benjamin2003} instead of the WISE survey, since GLIMPSE covers the field of J1809 and offers better sensitivity.
If a source has more than one MW counterpart located within $2''$, we consider the nearest counterpart to the X-ray source to be its MW counterpart.

To correct for the differences between the Spitzer/IRAC fluxes (3.6mag and 4.5mag) from the GLIMPSE survey and the WISE survey fluxes (W1 at 3.4 $\mu$m and W2 at 4.6 $\mu$m), that are used in the TD, we converted the IRAC magnitudes into WISE magnitudes using a linear regression model for sources that are detected in both GLIMPSE and WISE survey catalogs.

To compensate for the different definitions between the 3XMM-DR8 and the {\sl Chandra} energy bands outlined above in Section \ref{srcdet}, the 3XMM-DR8 fluxes are converted to the energy bands used here using scaling factors calculated by assuming a flat (in $\nu F_{\nu}$) spectrum (i.e.,  a photon index $\Gamma=2.0$). 
To clean the TD, several criteria are applied to filter out unreliable sources, including sources with large positional uncertainties, YSOs or stars with no MW counterparts (possibly due to large proper motions), and MW counterparts matched with isolated NSs  (by chance coincidence). 

Most of the AGNs in our TD are located off of the Galactic plane, where the Hydrogen column density and absorption are much lower than in the Galactic plane. 
Therefore, in order to compensate for this bias, all AGNs in the TD are reddened using the total dust extinction $E(B-V) = 20.46$ \citep{Schlegel1998} and the Galactic HI column density ($N_{\rm H}=1.53 \times 10^{22} {\rm cm}^{-2}$, \citealt{Dickey1990}) in the direction of the field of J1809. 
After applying the reddening, we remove any MW magnitudes of AGNs that are pushed beyond the detection limits of those MW surveys that are used.

\begin{table}
\caption{List of MW Features Used for ML Classification.}
\begin{center}
\begin{tabular}{c c}
\hline
Feature & Description\\
\hline
EP072Flux & Observed X-ray flux in the 0.7$-$2 keV band\\
EP27Flux & Observed X-ray flux in the 2$-$7 keV band\\
HR2 & Soft band hardness ratio  \\
HR4 & Hard band hardness ratio\\
Gmag & Gaia DR2 G band Magnitude \\
BPmag & Gaia DR2 BP band Magnitude\\
RPmag & Gaia DR2 RP band Magnitude\\
Jmag & 2MASS J band Magnitude \\
Hmag & 2MASS H band Magnitude \\
Kmag & 2MASS K band Magnitude \\
W1mag & WISE W1 Magnitude \\
W2mag & WISE W2 Magnitude \\
GR & Gmag $-$ RPmag\\
GB & Gmag $-$ BPmag\\
RB & RPmag $-$ BPmag \\
RW2 & RPmag $-$ W2mag \\
JH & Jmag $-$ Hmag\\
JK & Jmag $-$ Kmag\\
HK & Hmag $-$ Kmag \\
W1W2 & W1mag $-$ W2mag \\
\hline
\end{tabular}
\tablecomments{}{The catalogs used for the training dataset are listed in Section \ref{srcdet}. Since some of the catalogs used for the J1809 field differed from those used for the training dataset, the fluxes were converted to be compatible with the training dataset (see Appendix). }
\end{center}
\label{MWfeat}
\end{table}

All features that are used in the TD are listed in Table ~\ref{MWfeat}.
The fluxes in all bands (except the EP072Flux)  are divided by the $0.7$-$2$ keV flux to help mitigate the impact of their  unknown distances.  
Before passing the TD into the RF classifier, we standardize our data in the following way:
\begin{equation}
X=\frac{x_i-\mu_i}{\sigma_i}
\end{equation}
\noindent where $x_i$ is the $i$th feature of a source while $\mu_i$ and $\sigma_i$ are the corresponding feature's mean and standard deviation across the entire TD. 

To handle the large imbalance of source types (e.g., there are substantially more X-ray detected AGNs than neutron stars), we use an implementation of the Synthetic Minority Over-sampling Technique (SMOTE; \citealt{Chawla2011}) written in python\footnote{see \url{https://imbalanced-learn.readthedocs.io/en/stable/generated/imblearn.over\_sampling.SMOTE.html}} to oversample our training data. 
As for the missing data, which can either occur due to a source not being covered by a specific survey (which we try to mitigate by using all-sky surveys) or because the survey does not go deep enough to detect the MW counterpart to the source, we replace missing values with a large negative flag value of $-100$. 

\begin{figure*}
\begin{center}
\includegraphics[scale=0.5,trim=0 0 0 0,angle=0]{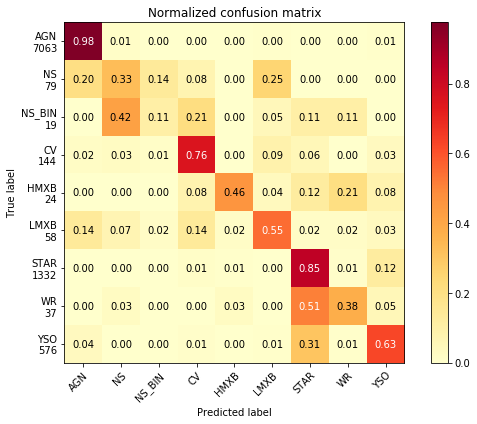}
\includegraphics[scale=0.5,trim=0 0 0 0,angle=0]{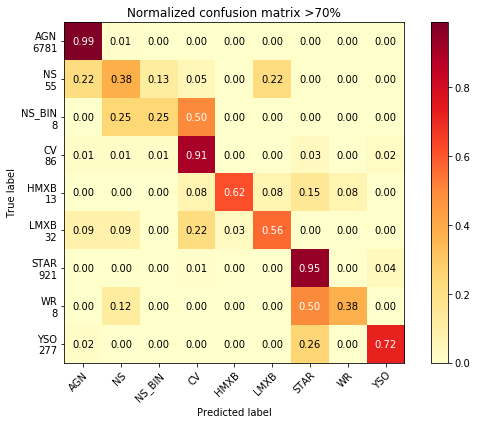}
\caption{The confusion matrices based on the TD (after splitting the TD in half; see Appendix). The left panel is for sources with all classification confidences and the right panel for sources with classification confidences $>70\%$. The total number of sources for each class are also shown under their true (vertical dimension) class. The predicted class is shown at the bottom (horizontal dimension).} 
\label{fig:confusion}
\end{center}
\end{figure*}

To check the accuracy of the algorithm, we randomly split the TD (after applying reddening on AGNs) into separate halves. 
Then we over-sample one half of the data with SMOTE and train RF classifier on this half of TD. 
The other half of the TD is then used to test the model. 
Then we reverse the two halves and do the evaluation again and use the average classification accuracies in the confusion matrices (i.e., also known as 2-fold cross-validation). 
The overall accuracy on the test set is $\sim92\%$ for all sources and $\sim97\%$ for sources with $>70\%$ classification confidences.  
However, we note that the total accuracy is being primarily driven by AGNs and stars which dominate the TD.
The normalized confusion matrices for all sources and for sources with ``confident'' classifications are shown in Figure \ref{fig:confusion}. 

\end{document}